\documentclass[preprint]{emulateapj}
\usepackage{hyperref}
\usepackage[usenames,dvipsnames]{color}
\hypersetup{colorlinks,citecolor=Blue,linkcolor=Red,urlcolor=Blue}

\usepackage[titletoc]{appendix}
\usepackage{amsmath}

\newcommand{\be}{\begin{equation}}
\newcommand{\ee}{\end{equation}}
\def\lta{\,\raise 0.3 ex\hbox{$ < $}\kern -0.75 em
 \lower 0.7 ex\hbox{$\sim$}\,}
\def\gta{\,\raise 0.3 ex\hbox{$ > $}\kern -0.75 em
 \lower 0.7 ex\hbox{$\sim$}\,} 
 
\newcommand{\rplan}{ R_{\rm p}}

\begin{document} 

\title{Resonant Removal of Exomoons during Planetary Migration} 

\author{Christopher Spalding$^1$, Konstantin Batygin$^1$, 
and Fred C. Adams$^{2,3}$} 
\affil{$^1$Division of Geological and Planetary Sciences\\
California Institute of Technology, Pasadena, CA 91125} 
\affil{$^2$Physics Department, University of Michigan, Ann Arbor, MI 48109} 
\affil{$^3$Astronomy Department, University of Michigan, Ann Arbor, MI 48109} 

\begin{abstract}
 Jupiter and Saturn play host to an impressive array of satellites, making it reasonable to suspect that similar systems of moons might exist around giant extrasolar planets. Furthermore, a significant population of such planets is known to reside at distances of several Astronomical Units (AU), leading to speculation that some moons thereof might support liquid water on their surfaces. However, giant planets are thought to undergo inward migration within their natal protoplanetary disks, suggesting that gas giants currently occupying their host star's habitable zone formed further out. Here we show that when a moon-hosting planet undergoes inward migration, dynamical interactions may naturally destroy the moon through capture into a so-called ``evection resonance." Within this resonance, the lunar orbit's eccentricity grows until the moon eventually collides with the planet. Our work suggests that moons orbiting within about $\sim10$ planetary radii are susceptible to this mechanism, with the exact number dependent upon the planetary mass, oblateness and physical size. Whether moons survive or not is critically related to where the planet began its inward migration as well as the character of inter-lunar perturbations. For example, a Jupiter-like planet currently residing at 1\,AU could lose moons if it formed beyond $\sim5$\,AU. Cumulatively, we suggest that an observational census of exomoons could potentially inform us on the extent of inward planetary migration, for which no reliable observational proxy currently exists.
\end{abstract}

\section{Introduction} 
\label{sec:intro} 

The past two decades have brought thousands of extrasolar planetary candidates to light. These systems have repeatedly challenged the notion that our Solar System is somehow typical \citep{Winn2015}. Notable examples include the existence of hot Jupiters \citep{Mayor1995}, spin-orbit misalignments \citep{Winn2010}, and the prevalence of highly compact, multi-planet systems \citep{Lissauer2011,Rowe2015}. However, as of yet, we have not been able to place the many known solar-system moons into their appropriate Galactic context. Observational surveys are now underway with this specific goal \citep{Kipping2009,Kipping2009b,Kipping2012,Kipping2015}. Motivated by the potential for upcoming exo-lunar detections, this work explores how the present-day configurations of exomoons might have been sculpted by dynamical interactions playing out during the epoch of planet formation.

Not long after the first detections of giant extrasolar planets \citep{Mayor1995}, speculations arose regarding what types of moons these bodies may host \citep{Williams1997}. Much of the interest has been astrobiological in nature - if giant planets reside in the habitable zones of their host stars, perhaps the moons thereof capable of sustaining liquid water on their surfaces \citep{Heller2014}. In contrast, any putative liquid water within the moons of Jupiter and Saturn must be maintained by way of tidal heating. Within the current observational dataset, however, our Solar System's giant planet configuration is by no means universal. A significant population of giant planets is found to reside between $\sim1-5$\,AU \citep{Dawson2013}, just inside the orbit of Jupiter\footnote{The apparent scarcity of planets at Jupiter's distance is subject to observational biases, not least owing to the associated long orbital periods. Recent searches are uncovering more distant bodies (e.g., \citealt{Knutson2014}).} (5.2\,AU).

Moons are expected to arise as an intrinsic outcome of giant-planet formation \citep{Canup2002,Mosqueira2010}. In particular, the core accretion model dictates that cores comprising multiple Earth masses of material, form outside their natal disks' ice line, before initiating a period of runaway gas accretion \citep{Pollack1996,Lambrechts2012}. Restricting attention to planets with masses greater than Saturn, their gravitational influence upon the protoplanetary disk will eventually clear a ``gap" in the gas within their vicinity \citep{Crida2006}. Material is capable of flowing through the gap and entering the planet's Hill sphere ($r_{\textrm{H}}$; the region around the planet where its potential dominates the motion of test particles). The residual angular momentum of the material is then distributed into a circumplanetary disk, extending out to $\sim0.4\,r_{\textrm{H}}$ \citep{Martin2011}, where moons are thought to form.

The angular momentum exchange associated with gap-clearing, in concert with viscous accretion within the protoplanetary disk, is expected to drive Type II migration of young planets, taking them to shorter-period orbits. Traditional theoretical treatments have suggested that the Type II migration rate is similar to the accretional velocity of disk gas \citep{Armitage2010,Kley2012}, though reality is likely more complicated \citep{Duffell2014}. Regardless, it is widely suspected that migration rates can be sufficient to reduce planetary semi-major axes by well over an order of magnitude within a typical disk lifetime (1-10\,Myr; \citealt{Haisch2001}, see below). Consequently, the `Jupiters' currently residing at several AU probably formed at more distant radii. Crucially, however, there currently exists no reliable, observational proxy that constrains the extent of migration.

 In this paper, we demonstrate that if the migrating planet hosts a moon, inward migration can lead to the moon's destruction by way of the ``evection resonance" (\citealt{Yoder1976,Touma1998,Cuk2012}). To illustrate the problem, consider the apsidal precession of a lunar orbit around an oblate giant planet. At large heliocentric distances, this precession is more rapid than the planetary mean motion about the central star. As the planet migrates inwards, its orbital frequency increases before becoming approximately commensurate with the lunar precession frequency. Assuming resonant capture (see section~[\ref{capture}]), further migration will pump the moon's eccentricity upwards until its pericenter approaches the planet's surface and the moon is lost.

 Our treatment here remains largely outside of the realm of hot Jupiters (giant planets with orbital periods of several days), whose reduced Hill spheres permit satellites only within a few planetary radii \citep{Domingos2006,Kipping2009}. Planetary migration may therefore remove moons around these objects without the resonant mechanism proposed here. Additionally, tidal planet-moon interactions further reduce the stability region of lunar orbits by expelling (or destroying) larger moons over Gyr timescales \citep{Barnes2002}. These issues make it difficult to relate current exolunar architectures of closer-in planets to their formation conditions. Consequently, we restrict our attention to bodies outside of $\sim0.5$\,AU from their stars.

The mechanism described herein requires both that the planet-moon system both begins outside of resonance and that migration proceeds until the moon is lost by collision. These conditions are quantified in section~(\ref{subseq}). The dynamics are critically dependent on the planetary radius planet and second gravitational moment $J_2$ (these determine the lunar precession frequency). Accordingly, we must begin with a brief discussion of reasonable parameters associated with young giant planets. 
 
 \subsection{Properties of young giant planets}\label{GiantPlanets}

Early models of giant planets naturally focussed on older planets, such as Jupiter and Saturn. The advantage here was that interior models lost their sensitivity to initial conditions over the relatively short ($\sim20$\,Myr) Kelvin-Helmholtz timescale \citep{Stevenson1982,Marley2007}. However, during the epoch of disk-driven planetary migration, the initial condition is crucial. Models extracting initial conditions from core accretion theory infer much smaller planetary radii $R_{\textrm{p}}$ than so-called "hot start" models, such as gravitational instability \citep{Marley2007}. For illustration, we focus on planets arising from core accretion, where radii sit close to $1.2-1.4$ times Jupiter's current radius $R_{\textrm{J}}$, but all further arguments could easily be applied to larger, hot-start planets. For the sake of definiteness, we choose $R_{\textrm{p}}=1.4\,R_{\textrm{J}}$ for the moon-hosting planet throughout this work.

 Eccentricity growth will remove moons either through physical collision with the planet or through tidal disruption, whichever happens earlier. Tidal disruption will occur close to the Roche radius (e.g., \citealt{Canup2010}) which, expressed in terms of satellite mass $m_{\textrm{s}}$, satellite radius $R_{\textrm{s}}$ and planetary mass $M_{\textrm{p}}$ may be written
 \begin{align}
 \frac{R_{\textrm{L}}}{R_{\textrm{p}}}\approx2.5 \bigg(\frac{M_{\textrm{p}}}{m_{\textrm{s}}}\bigg)^{\frac{1}{3}}\frac{R_{\textrm{s}}}{R_{\textrm{p}}}.
 \end{align}

 For parameters typical of Io-like bodies around young Jupiters ($R_{\textrm{p}}=1.4R_{\textrm{J}}$), $R_{\textrm{L}}/R_{\textrm{p}}<1$ and so moons are only lost by way of direct collision with the planet. Therefore, we consider a moon as lost when its pericenter approaches $R_{\textrm{p}}$ with the caveat that the Roche radii of more massive, compact planets may   indeed lie outside the planetary surface. 

 In addition to the planetary radius, an approximation for $J_2$ is required. For purely rotational deformation, the relationship between $J_2$, the Love number $k_2$ (twice the apsidal motion constant) and the planetary spin rate $\Omega$ may be expressed as (\citealt{Sterne1939}): 
 \begin{align}\label{J2}
 J_2=\frac{1}{3}\bigg(\frac{\Omega}{\Omega_{\textrm{b}}}\bigg)^2k_2,
 \end{align}
 where $\Omega_\textrm{b}^2\equiv GM_{\textrm{p}}/R_{\textrm{p}}^3$ is the break-up spin rate. Unfortunately, the above expression merely expresses one unknown quantity $J_2$ as a function of two other unknown quantities. However, $k_2$ can be estimated by modelling the planet as a polytrope with index $\chi=3/2$, \citep{Chandrasekhar1957,Batygin2013} yielding a Love number $k_2\approx0.28$. 
 
 It is more difficult to speculate upon $\Omega/\Omega_{\textrm{b}}$. The young giant planet in $\beta$ Pictoris\,b has had its spin period estimated at $\sim 8$\,hours \citep{Snellen2014}, close to what one would expect by extrapolating the equatorial velocities of the Solar-System's planets to the mass of $\beta$ Pictoris\,b (about $8M_{\textrm{J}}$). This result tentatively suggests that spin rates of young giant planets are little altered between 10\,Myr and 4.5\,Gyr after their formation, but the spin rate within the first 1\,Myr remains purely speculative. For the sake of definiteness, we take $J_2=0.02$ as a nominal value for young giant planets, slightly larger than Jupiter's current $J_2\approx0.015$ \citep{Murray1999}. We note, however, that $J_2$ may reasonable lie within the range $0.01>J_2>0.1$, with the upper bound deduced from equation~(\ref{J2}), and so further research is required to better constrain this quantity.
  
  \section{Evection Resonance} \label{resonance}

In this section, we quantitatively describe the dynamical influences upon a lunar orbit hosted by a young, giant planet. Consider the moon's orbit to have eccentricity $e$, inclination $i$ and semi-major axis $a_{\textrm{m}}$. The effect of planetary oblateness, $J_2$ is to force a precession of the longitude of pericenter $\varpi$ with frequency (for a derivation, see e.g. \citealt{Danby1992})
\begin{align}\label{omegadot}
\dot{\varpi}=\nu_{J_2}\approx \,\frac{3}{2}J_2\bigg(\frac{R_{\textrm{p}}}{a_{\textrm{m}}}\bigg)^2n_{\textrm{m}}\frac{1}{(1-e^2)^2}\bigg(2-\frac{5}{2}\sin^2(i)\bigg),
\end{align}
\noindent where $n_{\textrm{m}}$ is the mean motion of the lunar orbit and $R_\textrm{p}$ is the planetary radius. Note that the magnitude of $\nu_{J_2}$ increases monotonically with eccentricity, but its sign changes at a critical inclination of ($i_{\textrm{crit}}\approx63.4^\circ$). 

For simplicity, in all further analyses we will assume that the lunar orbit is coplanar with the planet's equator (that is, we set $i=0$) and, furthermore, that the planet itself has zero obliquity. These assumptions are motivated by the expectation that young giant planets inherit sufficient angular momentum from their natal disks to align both their spin axes and circumplanetary disks with their heliocentric orbits. It should be noted however that spin-orbit resonances have been proposed as an explanation for Saturn's obliquity \citep{Ward2004} and so similar dynamical processes may generate obliquities in moon-hosting planets. For the purposes of this work, we simply note that mild non-coplanarity slows the lunar precession rate which, as discussed below, leads to a more distant encounter with the evection resonance.

\begin{figure}
\centering
\includegraphics[trim=1cm 2.6cm 2cm 3cm, clip=true,width=1\columnwidth]{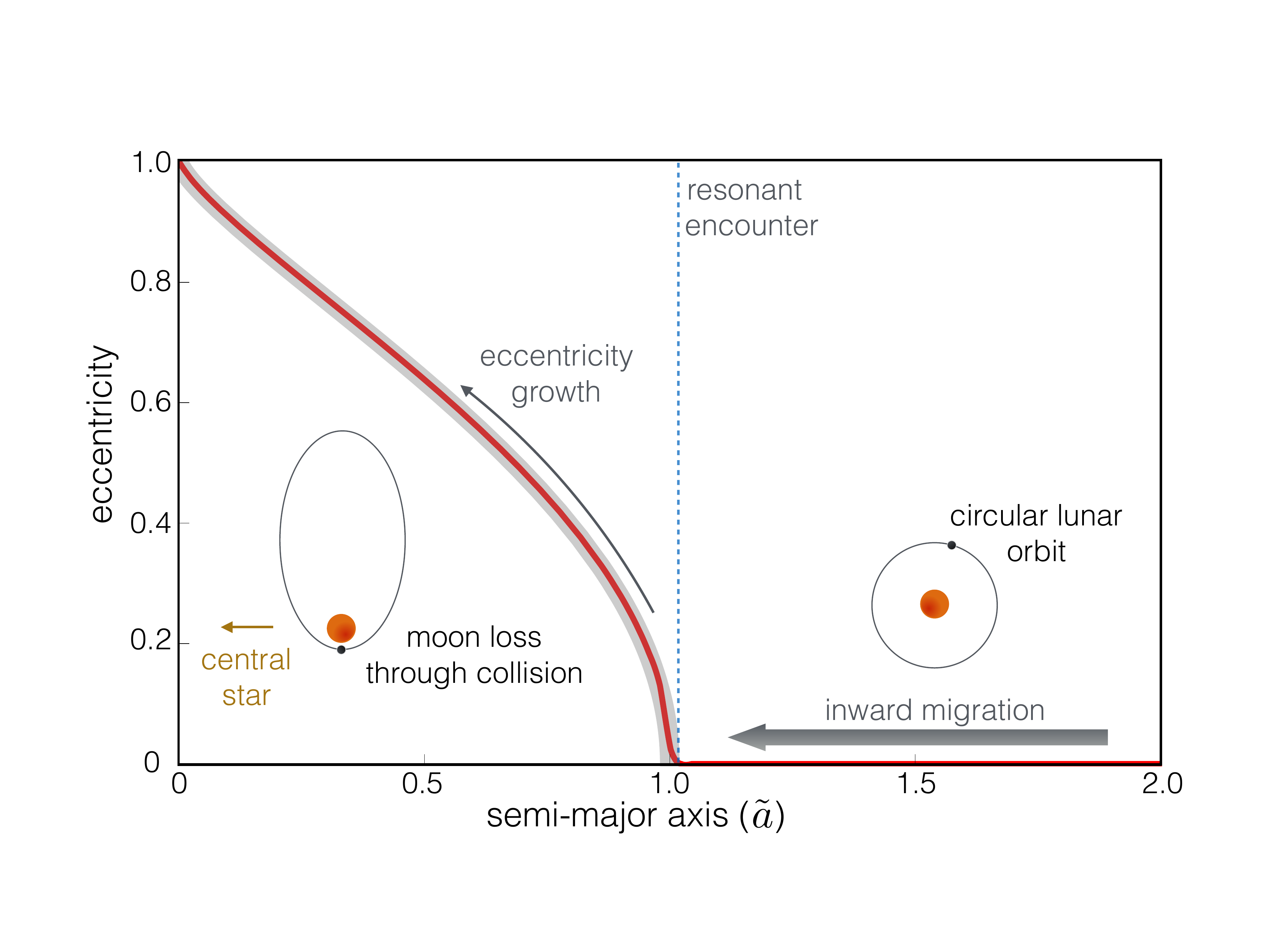}
\caption{Dimensionless illustration of resonant capture and eccentricity growth. The red line denotes the lunar eccentricity corresponding to the stable equilibrium of the Hamiltonian~\ref{Ham1}. The thicker grey line denotes the analytical expression~(\ref{Analytic}) describing exact resonance. The two solutions are almost indistinguishable.}
\label{EquE}
\end{figure}

 Provided the planet forms sufficiently far out, the precession frequency of the exomoon orbit will exceed the planetary mean motion $n_{\textrm{p}}$. During inward migration, $n_{\textrm{p}}$ increases until, at some point, the two frequencies $\nu_{J_2}$ and $n_{\textrm{p}}$ are approximately equal (Figure~\ref{EquE}), known as the evection resonance. This condition may be written in the form
\begin{align}\label{res}
{3\over2} J_2 \left( {R_{\textrm{p}} \over a_{\textrm{m}}} \right)^2 
\left({G M_{\textrm{p}} \over a_{\textrm{m}}^3}\right)^{1/2}\frac{1}{(1-e^2)^2} = 
\left({G M_\star \over a_{\textrm{p}}^3}\right)^{1/2} \,, 
\end{align}
 \noindent where $M_{\textrm{p}}$ is the mass of the planet, $a_{\textrm{p}}$ is the planetary orbital semi-major axis and $M_\star$ is the mass of the central star. Therefore, supposing the moon to originate at low eccentricity ($e\approx0$), resonance-crossing occurs at the heliocentric distance,
 \begin{align}\label{Circle}
 a_{\textrm{res}}=R_{\textrm{p}}\bigg[\bigg(\frac{2}{3J_2}\bigg)^2\bigg(\frac{a_{\textrm{m}}}{R_{\textrm{p}}}\bigg)^7\bigg(\frac{M_\star}{M_{\textrm{p}}}\bigg)\bigg]^{1/3}.
 \end{align}
 If the moon is caught into resonance, subsequent planetary migration drives the moon's orbital eccentricity to ever higher values. The physical source of eccentricity modulation is the torque supplied by the central star \citep{Touma1998,Cuk2012}. 

 In order to demonstrate the relevance of resonant capture under typical parameters, consider the planetary period $T_{\textrm{p}}$ corresponding the resonant condition above, 
\begin{align}
T_{\textrm{p}}\bigg|_{\textrm{res}}\approx 2700 \,\textrm{days} \bigg(\frac{J_2}{0.015}\bigg)^{-1}\bigg(\frac{a_{\textrm{m}}/R_{\rm{p}}}{a_{\textrm{Io}}/R_{\textrm{J}}}\bigg)^{\frac{7}{2}}.
\end{align}
We have scaled the parameters appropriately for the current Jupiter-Io configuration. Jupiter's orbital period is 4,332 days, meaning that, were Jupiter to be slowly forced in toward the Sun (and we ignore the influence of the other Jovian satellites), Io would encounter the evection resonance at roughly 3.8\,AU. Abundant extrasolar giant planets have thus far been detected with similar heliocentric distances ($\sim1-5$\,AU; \citealt{Dawson2013}), suggesting that the conditions for evection resonance might frequently be encountered in young giant planet-moon systems.

\subsection{The Evection Hamiltonian}
Criterion~(\ref{res}), describing an encounter with resonance takes a simple form, however, there is in general no guarantee that the moon will become captured into the resonance. Furthermore, assuming capture occurs, the subsequent evolution of eccentricity is non-trivial to compute. In order to tackle these aspects, we adopt a Hamiltonian approach, describing the lunar dynamics in terms of the combined gravitational potential of the central star and the planetary quadrupole ($J_2$). This section focuses on the dynamics of capture into resonance. The reader may skip to section~\ref{subseq} for a discussion of the dynamical loss of moons assuming capture occurs.

The Hamiltonian describing the dynamics of a moon in orbit around an oblate planet has been derived elsewhere (e.g. \citealt{Touma1994,Touma1998}). Despite their intuitive convenience, Keplerian orbital elements do not comprise a canonical set of coordinates. Accordingly, in order to utilize a symplectic form, we work in terms of reduced Poincar\'e (or, modified Delauney; \citealt{Murray1999,Morbidelli2002}) variables defined as follows:

\begin{align}\label{Poincare}
\Lambda\equiv m \sqrt{G\,M_{\textrm{p}}\,a}\,\,\,\,\,\,\,\,\,\,&\lambda\equiv M+\varpi+\Omega\nonumber\\
\Gamma\equiv \Lambda (1-\sqrt{1-e^2})\,\,\,\,\,\,\,\,\,\,&\gamma\equiv-\varpi-\Omega,
\end{align}

\noindent where $\Omega$ is the longitude of ascending node (not used owing to the assumption of coplanarity), $M$ is mean anomaly and subscripts `m' and `p' are used below to refer to the moon and the planet respectively. 
Physically, $\Lambda_{\textrm{m}}$ corresponds to the angular momentum the moon would possess on a circular orbit of semi-major axis $a_{\textrm{m}}$, and $\Gamma_{\textrm{m}}$ describes the angular momentum difference between the moon's true orbit and a circular orbit sharing its semi-major axis. We assume that the lunar orbital frequency is large enough to utilise a secular approach, whereby the Hamiltonian is ``averaged" over a lunar orbit. Consequently, explicit dependence upon $M$ is removed, extracting $\Lambda_{\textrm{m}}$ as an integral of the motion. In terms of the variables~(\ref{Poincare}), the Hamiltonian takes the form \citep{Touma1998}
 
 \begin{align}\label{Ham}
\mathcal{H}=&-\frac{1}{2}n_{\textrm{m}}\,J_2\bigg(\frac{R_{\textrm{p}}}{a_{\textrm{m}}}\bigg)^2\,\Lambda_{\textrm{m}}\bigg(\frac{\Lambda_{\textrm{m}}-\Gamma_{\textrm{m}}}{\Lambda_{\textrm{m}}}\bigg)^{-3}\nonumber\\
&-\frac{15}{8}\frac{n_{\textrm{p}}^2}{n_{\textrm{m}}}\Lambda_{\textrm{m}}\,\frac{\Gamma_{\textrm{m}}}{\Lambda_{\textrm{m}}}\bigg(2-\frac{\Gamma_{\textrm{m}}}{\Lambda_{\textrm{m}}}\bigg)\cos\big[2(n_{\textrm{p}}t+\gamma_{\textrm{m}})\big],
\end{align}
where the planet-moon system orbits the host star with mean motion $n_{\textrm{p}}$, such that $\lambda_{\textrm{p}}=n_{\textrm{p}} t$.

 The dynamics are best analyzed in a frame co-orbiting with the planet. Accordingly, we perform a canonical transformation of the above Hamiltonian using the new angle
\begin{align}\label{transform}
\tilde{\gamma}\equiv n_{\textrm{p}} t+\gamma
\end{align}
to obtain the autonomous Hamiltonian
 \begin{align}\label{Ham1}
\mathcal{H}=&n_{\textrm{p}} \Gamma_{\textrm{m}} -\frac{1}{2}n_{\textrm{m}}\,J_2\bigg(\frac{R_{\textrm{p}}}{a_{\textrm{m}}}\bigg)^2\,\Lambda_{\textrm{m}}\bigg(\frac{\Lambda_{\textrm{m}}-\Gamma_{\textrm{m}}}{\Lambda_{\textrm{m}}}\bigg)^{-3}\nonumber\\
&-\frac{n_{\textrm{p}}^2}{n_{\textrm{m}}}\Lambda_{\textrm{m}}\frac{15}{8}\,\frac{\Gamma_{\textrm{m}}}{\Lambda_{\textrm{m}}}\bigg(2-\frac{\Gamma_{\textrm{m}}}{\Lambda_{\textrm{m}}}\bigg)\cos(2\tilde{\gamma}_{\textrm{m}}).
\end{align}
 \noindent The first term arises as a result of transformation~(\ref{transform}), the second term describes the influence of planetary oblateness upon the lunar orbit, and gives rise to the precession frequency~\ref{omegadot}. The third term is new and describes the secular perturbation upon the moon's orbit raising from the star. Note that the Gaussian averaging process is inertially equivalent to considering the orbit of the moon to act as an eccentric, massive wire. Thus, the third term arises from the torques communicated between the stellar gravitational potential and an eccentric wire. 
 
It is appropriate to scale the action $\Gamma_{\textrm{m}}$ by the integral of motion $\Lambda_{\textrm{m}}$, thus defining a new canonical momentum
 \begin{align}
\tilde{\Gamma}_{\textrm{m}}\equiv\frac{\Gamma_{\textrm{m}}}{\Lambda_{\textrm{m}}}.
 \end{align}
 \noindent In order to preserve symplectic structure, we likewise scale the Hamiltonian itself by $\Lambda_{\textrm{m}}$, yielding
  \begin{align}\label{Ham}
\tilde{\mathcal{H}}=&n_{\textrm{p}} \tilde{\Gamma}_{\textrm{m}} -\frac{1}{2}n_{\textrm{m}}\,J_2\bigg(\frac{R_{\textrm{p}}}{a_{\textrm{m}}}\bigg)^2\,\big(1-\tilde{\Gamma}_{\textrm{m}}\big)^{-3}\nonumber\\
&-\frac{n_{\textrm{p}}^2}{n_{\textrm{m}}}\frac{15}{8}\,\tilde{\Gamma}_{\textrm{m}}\big(2-\tilde{\Gamma}_{\textrm{m}}\big)\cos(2\tilde{\gamma}_{\textrm{m}}),
\end{align}
\noindent such that the system evolves according to Hamilton's equations in the form
\begin{align}\label{timee}
\frac{d\tilde{\gamma}_{\textrm{m}}}{dt}&=\frac{\partial \tilde{\mathcal{H}}}{\partial\, \tilde{\Gamma}_{\textrm{m}}}\nonumber\\
\frac{d\,\tilde{\Gamma}_{\textrm{m}}}{dt}&=-\frac{\partial \tilde{\mathcal{H}}}{\partial \tilde{\gamma}_{\textrm{m}}}.
\end{align}

As mentioned earlier, we consider inward planetary migration (increasing $n_{\textrm{p}}$), but do not explicitly consider the case where the moon itself is migrating within a circumplanetary disk \citep{Canup2002}. Qualitatively, the effect of inwards moon-migration would be to postpone the crossing of an evection resonance by increasing the influence of the planetary quadrupole. Additionally, we assume that any variations in the radius of the planet and its $J_2$ during the nebular epoch are negligible compared to the influence of variations in $n_{\textrm{p}}$. 

\subsection{Capture into resonance}\label{capture}

In this section, we outline the conditions under which moons are expected to become captured into resonance. The moon's orbital eccentricity is likely to be small during resonance passage and so we analyze the dynamics of capture using the small-$e$ (and, consequently, small-$\tilde{\Gamma}_{\textrm{m}}$) approximation to Hamiltonian~(\ref{Ham}) (e.g., \citealt{Touma1998}):
\begin{align}\label{Smalle}
\tilde{\mathcal{H}}&\approx \bigg[n_{\textrm{p}}-\frac{3}{2}n_{\textrm{m}} J_2 \bigg(\frac{R_{\textrm{p}}}{a_{\textrm{m}}}\bigg)^2\bigg]\tilde{\Gamma}_{\textrm{m}}-3\,n_{\textrm{m}} J_2 \bigg(\frac{R_{\textrm{p}}}{a_{\textrm{m}}}\bigg)^2 \tilde{\Gamma}_{\textrm{m}}^2\nonumber\\
&-\frac{15}{4}n_{\textrm{p}}\bigg(\frac{n_{\textrm{p}}}{n_{\textrm{m}}}\bigg)\tilde{\Gamma}_{\textrm{m}}\cos(2 \tilde{\gamma}_{\textrm{m}}).
\end{align}  

  \citet{Borderies1984} computed the probably for resonant capture of a system governed by the integrable single-parameter Hamiltonian
\begin{align}\label{Ham2}
\mathcal{H}'=-(1+2\delta)\Phi+2\Phi^2-\Phi \cos(2\phi),
\end{align}
 \noindent and so we make progress by casting Hamiltonian~(\ref{Smalle}) into a similar form.

 First, we scale both the Hamiltonian and the actions $\tilde{\Gamma}_{\textrm{m}}$ by a factor $\eta$ such that
 \begin{align}
\Phi=\frac{\tilde{\Gamma}_{\textrm{m}}}{\eta}\,\,\,\,\,\eta=\frac{5}{2}\frac{1}{J_2}\bigg(\frac{a_{\textrm{m}}}{\rplan}\bigg)^2\bigg(\frac{n_{\textrm{p}}}{n_{\textrm{m}}}\bigg)^2.
 \end{align}  
 \noindent This choice of $\eta$ ensures a common factor, 
 \begin{align}
  \nu'=\frac{15}{4}n_{\textrm{p}} \bigg(\frac{n_{\textrm{p}}}{n_{\textrm{m}}}\bigg)
  \end{align}
  \noindent between the coefficients of $2\Phi^2$ and $\Phi \cos(\phi)$. Dividing the Hamiltonian by this factor reproduces the form~(\ref{Ham2}), with the caveat that time must now be measured in units of $\nu'^{-1}$. That is, we have introduced a ``slow" canonical time
   \begin{align}
   \tau=\frac{15}{4}n_{\textrm{p}} \bigg(\frac{n_{\textrm{p}}}{n_{\textrm{m}}}\bigg)t.
   \end{align}
 By inspection, we see that the ``resonance proximity parameter"
 \begin{align}\label{delta}
 \delta=-\frac{1}{2}+\frac{2}{15}\frac{n_{\textrm{m}}}{n_{\textrm{p}}}-\frac{1}{5}J_2\bigg(\frac{\rplan}{a_{\textrm{m}}}\bigg)^2\bigg(\frac{n_{\textrm{m}}}{n_{\textrm{p}}}\bigg)^2,
 \end{align}
 which is highly negative for planetary orbits far outside of resonance (large $a_{\textrm{m}}$), but increases upon inward migration.
  
For dynamics governed by Hamiltonians of the form~(\ref{Ham2}), capture \textit{in the adiabatic limit} is certain for $\Phi<1/2$ \citep{Borderies1984}. In our case, this condition corresponds to a lunar eccentricity $e_{\textrm{cap}}$, above which, adiabatic capture is not guaranteed. Within the small $e$ approximation, $\tilde{\Gamma}_{\textrm{m}}\approx e^2/2$ and so we find the critical eccentricity, below which resonant locking is certain, to be 
 \begin{align}\label{ecap}
 e_{\textrm{cap}}=\sqrt{\frac{5}{2J_2}}\bigg(\frac{a_{\textrm{m}}}{\rplan}\bigg)\bigg(\frac{n_{\textrm{p}}}{n_{\textrm{m}}}\bigg).
 \end{align}
 \noindent At resonance-crossing, $n_{\textrm{p}}\approx(3/2)J_2(R_{\textrm{p}}/a_{\textrm{m}})n_{\textrm{m}}$ and so the criterion above yields the condition
 
 \begin{align}
  e&\lesssim\frac{3}{2}\sqrt{\frac{5}{2}}J_2^{\frac{1}{2}}\bigg(\frac{R_{\textrm{p}}}{a_{\textrm{m}}}\bigg)\nonumber\\
  &=0.03 \bigg(\frac{J_2}{0.02}\bigg)^{\frac{1}{2}}\bigg(\frac{R_{\textrm{p}}/a_{\textrm{m}}}{1/10}\bigg),
 \end{align}
 where we have chosen $R_{\textrm{p}}/a_{\textrm{m}}=1/10$ as a reference value because, as discussed later, more distant orbits are typically only lost outside of the adiabatic regime. 
 
 Note that $e=0.03$ is significantly larger than the eccentricities of the Galilean Satellites, but approaches that of Titan ($e=0.028$; \citealt{Iess2012}). Owing to their position within a circumplanetary disk, we expect that any young moons will possess eccentricities at least as small as the Galilean Satellites and ought therefore to be captured in the adiabatic regime. However, the presence of moon-moon resonances or other sources of eccentricity-pumping may quench the evection resonance in specific cases.

 \subsubsection{The adiabatic criterion}

It is well known in celestial mechanics that passing through resonances sufficiently rapidly can prevent capture \citep{Quillen2006} by way of leaving the `adiabatic regime.' Adiabatic motion occurs when the libration timescale of the moon within resonance is shorter than the timescale of resonance crossing. When satisfied, adiabatic motion allows the lunar orbit to grow in eccentricity, and therefore precession frequency, keeping pace with the rising planetary mean motion.

\begin{figure*}
\centering
\includegraphics[trim=0cm 13.2cm 1cm 0.1cm, clip=true,width=1\textwidth]{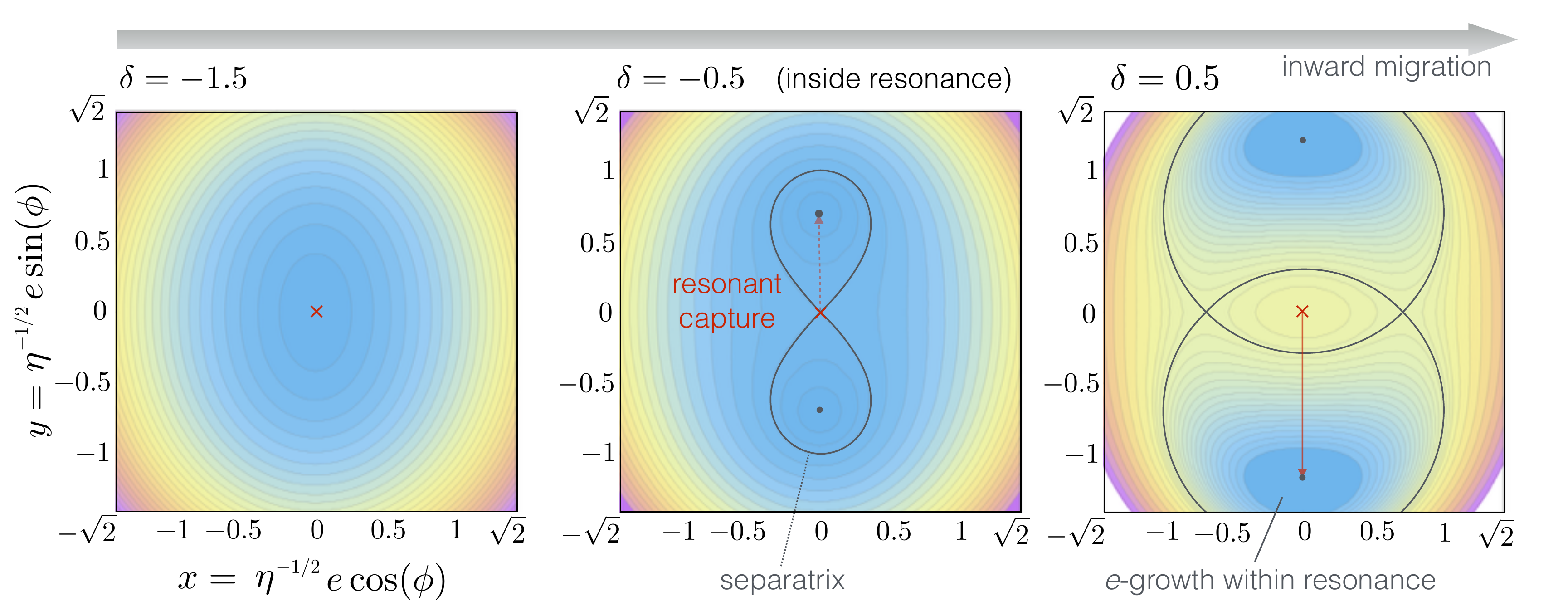}
\caption{Contours of the small-eccentricity Hamiltonian derived in the text under three scenarios. On the left, the moon is outside of resonance and all trajectories circulate about the origin. The middle panel represents the situation when the moon is inside resonance and the stable, null equilibrium eccentricity in the left panel has bifurcated into two, non-zero stable solutions and one unstable solution. On the right, we illustrate the situation once the moon leaves resonance. If capture occurs, the trajectory remains near the upper or lower equilibria. Unsuccessful capture causes the system to remain within the circulating region around the null-eccentricity equilibrium.}
\label{PhasePortrait}
\end{figure*}

 The adiabaticity criterion is best derived by changing to the canonical Cartesian coordinates,
\begin{align}\label{Cartesian}
x&=\sqrt{2\Phi}\cos(\phi)\propto e\cos(\phi)\nonumber\\
y&=\sqrt{2\Phi}\sin(\phi)\propto e \sin(\phi),
\end{align}
 where the proportionalities are valid in the small-$e$ limit. Performing the transformation, Hamiltonian~(\ref{Ham2}) takes the form
 \begin{align}\label{Ham3}
 \mathcal{H}=(1+2\delta)\bigg(\frac{x^2+y^2}{2}\bigg)-2\bigg(\frac{x^2+y^2}{2}\bigg)^2-\bigg(\frac{x^2-y^2}{2}\bigg).
 \end{align}
 \noindent  In Figure~\ref{PhasePortrait}, we plot contours of Hamiltonian~(\ref{Ham3}) for a range of values of $\delta$, where it can be seen that the number of equilibria increases from one to three to five upon increasing from $\delta=-1.5$ to $\delta=0.5$. We may compute when resonance is encountered by quantifying the fixed points of Hamiltonian~(\ref{Ham3}). On the $y$-axis ($\phi=\pi/2,3\pi/2$), fixed points occur at 
 \begin{align}
 y=0,\,\,\,y=\pm \sqrt{1+\delta}.
 \end{align}
 and so the equilibria away from $y=0$ exists for $\delta>-1$. As $\delta$ continues to grow, equilibria appear on the $x$-axis at 
 \begin{align}
 x=0,\,\,\,x=\pm\sqrt{\delta},
 \end{align}
 when $\delta\geqslant0$ (see Figure~\ref{PhasePortrait}). Accordingly, as the planet migrates inwards, resonance is encountered at $\delta=-1$ and an inner, circulation region develops at $\delta=0$.\footnote{The exact resonant position~(\ref{Circle}) corresponds to $\delta=-1/2$.} In other words, the `width' of the resonance is equivalent to $\Delta\delta=1$, corresponding to the amount of migration the planet must undergo to take its moon from outside to inside of resonance.
 
 A non-adiabatic crossing of resonance corresponds to the transitioning from outer to inner circulation in less than one oscillation period. This crossing time is given by
 \begin{align}
 \frac{d\delta}{dt}\approx\frac{\Delta \delta}{\Delta t}=\frac{1}{\Delta t}.
 \end{align}
\noindent We now suppose the planetary migration proceeds on the characteristic timescale $\tau_{\textrm{m}}$, such that
\begin{align}\label{adot}
\frac{1}{a_{\textrm{p}}}\frac{da_{\textrm{p}}}{dt}=-\frac{1}{\tau_{\textrm{m}}}.
\end{align}
\noindent With this prescription for $a_{\textrm{p}}$, we may take the time derivative of $\delta$ (Equation~(\ref{delta})) yielding the resonance crossing time,
\begin{align}
\Delta t=\frac{5 a_{\textrm{m}}^2n_{\textrm{p}}^2}{n_{\textrm{m}}(a_{\textrm{m}}^2n_{\textrm{p}}-3\rplan^2J_2n_{\textrm{m}})}\tau_{\textrm{m}}.
\end{align}
 
 For definiteness, we evaluate $\Delta t$ when $\delta=-1/2$, which is equivalent to condition given in equation~(\ref{resonance}). Adopting this midway point, the resonance crossing time is given by 
 
 \begin{align}
\Delta t=5\frac{n_{\textrm{p}}}{n_{\textrm{m}}}\tau_{\textrm{m}}.
\end{align}

All that remains is to estimate the libration timescale. Let us analyze the local neighborhood of $\mathcal{H}$ around the resonant fixed point ($x_{\textrm{eq}}=0$, $y_{\textrm{eq}}=\sqrt{1+\delta}$). Recalling that, at $x=0$, $\phi=\pi/2$ and $\Phi=\sqrt{2\,y}$, we define the variables
\begin{align}
\bar{\Phi}=\Phi-\frac{y_{\textrm{eq}}^2}{2}=\Phi-\frac{1+\delta}{2}\nonumber\\
\bar{\phi}=\phi-\frac{\pi}{2}
\end{align}
 which measure the distance away from the equilibrium fixed point. We now expand the Hamiltonian~(\ref{Ham3}) as a Taylor series to second order in $\bar{\Phi}$ and $\bar{\phi}$, setting $\delta=-1/2$. After one final scaling of the variables
 \begin{align}
 \hat{\Phi}=\frac{1}{\sqrt{2}}\bar{\Phi}\,\,\,\,\,\,\,\,\,\hat{\phi}=\frac{1}{\sqrt{2}}\bar{\phi},
 \end{align}
 
  \noindent we arrive at the local Hamiltonian
 \begin{align}
 \bar{\mathcal{H}}=-\frac{1}{2}\omega(\hat{\Phi}^2+\hat{\phi}^2),
 \end{align} 
 \noindent where the corresponding ``harmonic oscillator" frequency around this fixed point is $\omega=2$. We now convert back into real time units, and obtain the libration period  
 \begin{align}\label{libration}
 P_{\,\textrm{lib}}=2\pi(\omega \tau)^{-1}=\frac{4\pi}{15}\bigg(\frac{n_{\textrm{m}}}{n_{\textrm{p}}}\bigg)\frac{1}{n_{\textrm{p}}}.
 \end{align}
 \noindent Equating this quantity to the resonance crossing time, we arrive at the adiabatic criterion, expressed in terms of the planetary migration timescale: 
\begin{align}\label{criterion}
 \boxed{\frac{P_{\,\textrm{lib}}}{\Delta t}=\frac{2}{75}\bigg(\frac{n_{\textrm{m}}}{n_{\textrm{p}}}\bigg)^3\bigg(\frac{2\pi}{n_{\textrm{m}}}\bigg)\frac{1}{\tau_{\textrm{m}}}\lesssim1}
\end{align}

 We may immediately substitute in the resonance criterion~(\ref{res}) to determine the requirement for adiabatic migration in terms of lunar semi-major axis. The planetary migration timescale is likely to scale with planetary Keplerian orbital period $T_{\textrm{p}}$ \citep{Tanaka2002} and so it makes sense to likewise scale the adiabatic criterion: 
 \begin{align}\label{capture}
 \frac{\tau_{\textrm{m}}}{T_{\textrm{p}}}\gtrsim30\, \bigg(\frac{J_2}{0.02}\bigg)^{-2}\bigg(\frac{a_{\textrm{m}}}{R_{\textrm{p}}}\bigg)^4.
 \end{align}
 \noindent This dependence comes about because moons at larger $a_{\textrm{m}}/R_{\textrm{p}}$ are resonant at greater $a_{\textrm{p}}/a_{\textrm{m}}$, such that the typical libration timescales are reduced and the adiabatic criterion is easier to break. (As found above, more distant moons are also more likely to break the requirement $e<e_{\textrm{cap}}$.)
 \subsection{The Adiabaticity of Planetary Migration}
 
 In the above derivation, we supposed that the semi-major axis of the planet decays over a characteristic timescale $\tau_\textrm{m}$. The exact value of $\tau_\textrm{m}$, i.e., the rate of Type II migration, is still an active area of research \citep{Kley2012}. In this work, we adopt the reasonable, first order approximation that once giant planets open a gap in the protoplanetary disk, they migrate inwards with the accretionary flow (but see \citealt{Duffell2014}). Utilizing the Shakura-Sunyaev form for disk effective viscosity \citep{Shakura1973}, the accretionary velocity is given by \citep{Armitage2011}, 
 \begin{align}
v_{\textrm{acc}}\approx -\frac{3}{2}\alpha \bigg(\frac{h}{a_{\textrm{p}}}\bigg)^2 \Omega_K a_{\textrm{p}}
 \end{align}
 where $h$ is the scale height of the disk, $\alpha$ is the dimensionless turbulence parameter and $\Omega_K$ is the Keplerian velocity at radius $a_{\textrm{p}}$ in the disk. From this equation, we derive the form for the time evolution of the planetary semi-major axis, 
 \begin{align}
 \frac{1}{a_{\textrm{p}}}\frac{da_{\textrm{p}}}{dt}\approx-\frac{3}{2}\alpha \bigg(\frac{h}{a_{\textrm{p}}}\bigg)^2 \Omega_K ,
 \end{align}
 which we may estimate by supposing the disk aspect ratio $h/a_{\textrm{p}}\sim10^{-1}$. The value of $\alpha$ (and even the validity of its usage) is widely debated, and probably varies throughout the disk, depending upon which mechanisms dominate turbulent motions \citep{Hartmann1998,King2007,Armitage2011}. That said, the inferred value is usually within the range $10^{-4}<\alpha<10^{-2}$. Substituting these parameter values in for the migration timescale, we obtain reasonable bounds on the adiabaticity parameter  
 \begin{align}
 10^4\lesssim\tau_m/T_{\textrm{p}}\lesssim 10^6.
 \end{align} 
 Using the above criteria, we may now estimate the most distant exolunar orbit that is guaranteed to be captured. From condition~(\ref{capture}), 
 \begin{align}
 \bigg(\frac{a_{\textrm{m}}}{R_{\textrm{p}}}&\bigg)^4=\frac{1}{30}\frac{\tau_m}{T_{\textrm{p}}}\bigg(\frac{J_2}{0.02}\bigg)^{2},
 \end{align}
 we obtain the requirement for adiabatic capture that
 \begin{align}\label{capture4}
 \frac{a_{\textrm{m}}}{R_{\textrm{p}}}&\lesssim13\bigg(\frac{J_2}{0.02}\bigg)^{\frac{1}{2}}\bigg(\frac{\tau_m/T_{\textrm{p}}}{10^6}\bigg)^{\frac{1}{4}}.
 \end{align}
 To put this number into perspective, 13 planetary radii of $R_{\textrm{p}}=1.4 R_{\textrm{J}}$ sits outside of the current orbit of Ganymede (for which, $a_{\textrm{m}}/R_{\textrm{p}}\approx11$). Stable lunar orbits may exist out to roughly 1/3 to 1/2 of the Hill Radius \citep{Nesvorny2003}, meaning that a Jupiter-mass planet, residing beyond about 0.5\,AU from its host star, may possess moons too far out for adiabatic capture. Capture can still occur outside of the adiabatic limit, but the probability drops rapidly. Consequently, in the rest of the paper, we focus on moons situated at $a_{\textrm{m}}/R_{\textrm{p}}\lesssim10$, but maintain the caveat that specific cases may exist where capture occurs outside the regime of guaranteed capture.

 \section{Evolution within resonance}\label{subseq}
 
In this section, we calculate the evolution of the moon's eccentricity within resonance, assuming the planet-moon system satisfies the capture criteria given by equations~(\ref{ecap}) and (\ref{capture4}). Furthermore, we derive the conditions under which the lunar pericenter $a_{\textrm{m}}(1-e)$ coincides either with the Roche radius of its host planet, or the planetary radius itself. The Roche radii of young, Jupiter-mass planets are likely to reside inside the planetary radius (see section~[\ref{GiantPlanets}] above) and so we consider a planet-crossing orbital trajectory as the criterion for moon loss, which occurs at an eccentricity
\begin{align}
e_{\textrm{coll}}=1-\frac{R_{\textrm{p}}}{a_{\textrm{m}}}.
\end{align}

 We only consider the case whereby moons are lost at the planetary radius, but mention that higher-mass, compact gas giants might lose moons through tidal stripping, potentially generating a primordial ring system \citep{Canup2010}. Furthermore, planets forming under the ``hot start" regime, as opposed to core-accretion, will have significantly larger radii, lowering the required eccentricity for moon-loss \citep{Marley2007}.

\subsection{Lunar eccentricity growth}

\begin{figure}
\centering
\includegraphics[trim=3cm 6cm 4.9cm 3cm, clip=true,width=1\columnwidth]{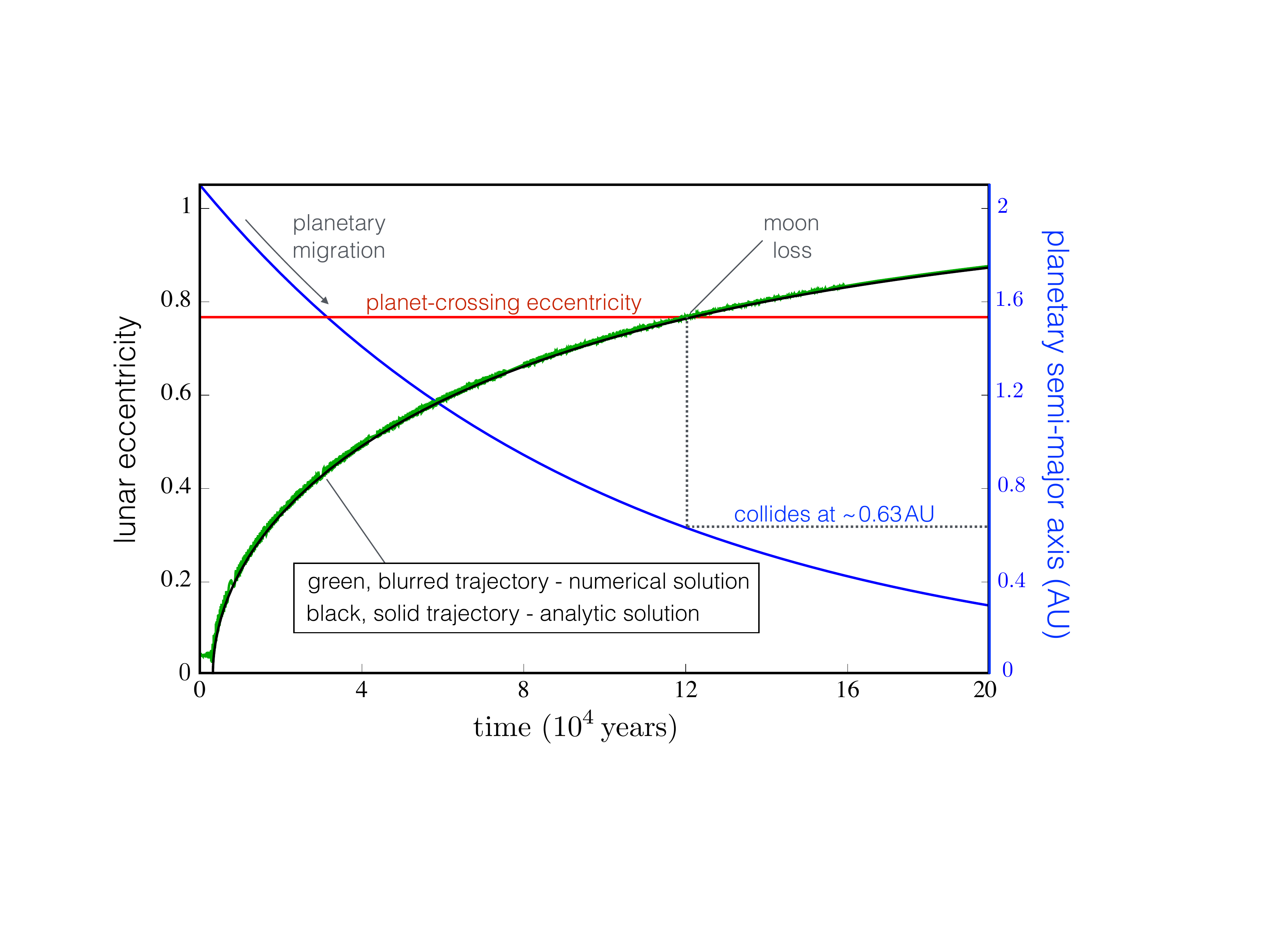}
\caption{Numerical solution of the capture and subsequent loss of a moon with semi-major axis equal to that of Io's current value. The green, jagged line follows the numerical solution where the black line close to its center illustrates the analytic solution derived in the text for the equilibrium of the the Hamiltonian (equation~(\ref{Analytic})). The horizontal, red line denotes the eccentricity at the which the lunar orbit crosses the planet's surface and the blue line, decreasing from left to right, depicts planetary migration. We consider parameters typical to the Io-Jupiter system except with the planetary radius inflated to $1.4R_{\textrm{J}}$. Notice that the analytic solution is almost indistinguishable from the mean lunar trajectory. The addition of dissipation collapses the numerical curve on top of the analytic one, provided the dissipation is not too severe (e.g., excessive tides, see Figure~\ref{TidesFigure}).}
\label{Numerical}
\end{figure}

As mentioned above, when the system crosses $\delta=-1$ from below (by way of inward migration), the single equilibrium at $e=0$ becomes unstable and undergoes a bifurcation into two stable equilibria appearing at non-zero eccentricities (Figures~[\ref{EquE},~\ref{PhasePortrait}]). Provided the lunar eccentricity begins relatively small, the resonant orbit will perform small-amplitude oscillations about the eccentricity corresponding to the equilibrium fixed point of the Hamiltonian. As long as the evolution proceeds within the adiabatic regime, quasi-conservation of phase-space area guarantees that the oscillation amplitude will remain small (see Figure~\ref{Numerical}). Moreover, dissipative processes, stemming from tides or disk interactions, will reduce the amplitude of these oscillations, causing the moon to very closely track the fixed points.

 Cumulatively, \textit{we may determine the evolution of the moon's eccentricity by solving for the stable fixed points of the governing Hamiltonian}. The small-$e$ case considered in the previous section was sufficient for analysis of adiabatic capture but we must work with unrestricted eccentricities in order to accurately describe evolution of the moon within resonance. The first-order approximation is to assume that the equilibrium eccentricity corresponds to an exact balance between the lunar precession period and planetary mean motion, described by equation~(\ref{res}). It is sensible to work in terms of a dimensionless semi-major axis 
\begin{align}
\tilde{a}\equiv\frac{a_{\textrm{p}}}{a_{\textrm{res}}},
\end{align}
such that, in solving the criterion~(\ref{res}) we find the moon's resonant eccentricity growth is well-described by
\begin{align}\label{Analytic}
\boxed{e_{\textrm{eq}}=\sqrt{1-\tilde{a}^{\frac{3}{4}}}}.
\end{align}
\noindent One can show, both perturbatively and through numerical solution, that the above expression corresponds very closely with the exact equilibrium of Hamiltonian~(\ref{Ham1}). Such an equilibrium may be obtained by a similar approach as was used above to calculate the small-$e$ equilibria, except that the resulting polynomial is non-trivial to solve.

We plot both the approximate solution~(\ref{Analytic}) and the exact, numerical solution in Figure~\ref{EquE} to demonstrate their similarity. Furthermore, in Figure~\ref{Numerical}, we compare the analytic expression~(\ref{Analytic}) (the black line) to a direct numerical solution of Hamilton's equations (the oscillating, green line), where it is apparent that the approximate solution is more than adequate to describe the eccentricity growth of the moon.

 \subsection{Condition for Moon-Loss}
 With an analytic solution for $e_{\textrm{eq}}$ in hand, we are now in a position to calculate the semi-major axis $a_{\textrm{coll}}$ at which a resonant moon will collide with the planet. We suppose the moon to be lost at $e_{\textrm{eq}}=e_{\textrm{coll}}$ which, from~(\ref{Analytic}) and~(\ref{collision}), occurs when 
 \begin{align}\label{collision}
 1-\frac{R_{\textrm{p}}}{a_{\textrm{m}}}=\sqrt{1-\bigg(\frac{a_{\textrm{coll}}\,\,}{a_{\textrm{res}}}\bigg)^{\frac{3}{4}}}.
 \end{align}
 \noindent For convenience, we define the dimensionless variable  
 \begin{align}
 r\equiv \frac{a_{\textrm{m}}}{R_{\textrm{p}}},
 \end{align} 
\noindent  such that the condition for a moon being lost (after subbing in for $a_{\textrm{res}}$) becomes
 \begin{align}\label{quartic}
 \bigg(1-\frac{1}{r_{\textrm{loss}}}\bigg)^{2}=1-\bigg(\frac{a_{\textrm{coll}}}{R_{\textrm{p}}}\bigg)^{\frac{3}{4}}r_{\textrm{loss}}^{-\frac{7}{4}}\bigg(\frac{M_\star}{M_{\textrm{p}}}\bigg)^{-\frac{1}{4}}J_2^{\frac{1}{2}}\bigg(\frac{2}{3}\bigg)^{-\frac{1}{2}},\nonumber\\
 \end{align}
\noindent where $r_{\textrm{loss}}$ is the dimensionless semi-major axis of a moon lost at heliocentric distance $a_{\textrm{p}}$ . An analytic solution exists for $r_{\textrm{loss}}$ above, but its functional form is rather complicated (though analytic approximations exist).

As mentioned above, in order to capture a moon into the evection resonance, the planet must originate outside of resonance ($a_0>a_{\textrm{res}}$, where $a_0$ is the location of the planet at the time of moon formation). This condition may be recast in terms of the the most distant lunar orbit that would encounter resonance ($r=r_{\textrm{Max}}$) as the planet migrates from $a_{\textrm{p}}=a_0$. Rearranging the expression for $a_{\textrm{res}}$, we obtain
 \begin{align}\label{max}
 r_{\textrm{Max}}&=\bigg[\bigg(\frac{3}{2}\bigg)^{2}\bigg(\frac{a_0}{R_{\textrm{p}}}\bigg)^{3}\bigg(\frac{M_{\textrm{p}}}{M_{\star}}\bigg)J_2^2\bigg]^{1/7}
 \end{align} 
 \noindent In other words, any moons forming further than $r_{\textrm{Max}}$ from their host planet will not become captured during subsequent inward migration. Combined with the condition for moon loss $r=r_{\textrm{loss}}$ above, we may define an ``exclusion zone", within which, moons may be lost via the evection resonance as a planet migrates from $a_0$ to $a_{\textrm{coll}}$:
\begin{align}\label{zone}
\boxed{r_{\textrm{loss}}<r_{\textrm{excl}}<r_{\textrm{Max}}},
\end{align}
\noindent where expressions for $r_{\textrm{loss}}$ and $r_{\textrm{Max}}$ are given by equations~(\ref{collision}) and~(\ref{max}).

Crucially, the excluded region's outer edge $r_{\textrm{Max}}$ depends only upon where the planet began its inward migration, $a_0$. Consequently, if such a region is observed among future exomoon detections, its outer edge may be used to directly constrain where the planet-moon system formed, irrespective of where the planet resides presently. This is fortunate, because giant planets have long been suspected to undergo planetesimal-driven migration following the epoch of disk-driven migration \citep{Tsiganis2005}. We note, additionally, that the extent of post-disk planetary migration may in principle be inferred from the discrepancy between the current planetary position and that derived from expression~(\ref{collision}) for $r_{\textrm{loss}}$.

\subsection{Illustrative example of exclusion zone}

\begin{figure*}
\centering
\includegraphics[trim=1cm 4.8cm 2cm 2cm, clip=true,width=1\textwidth]{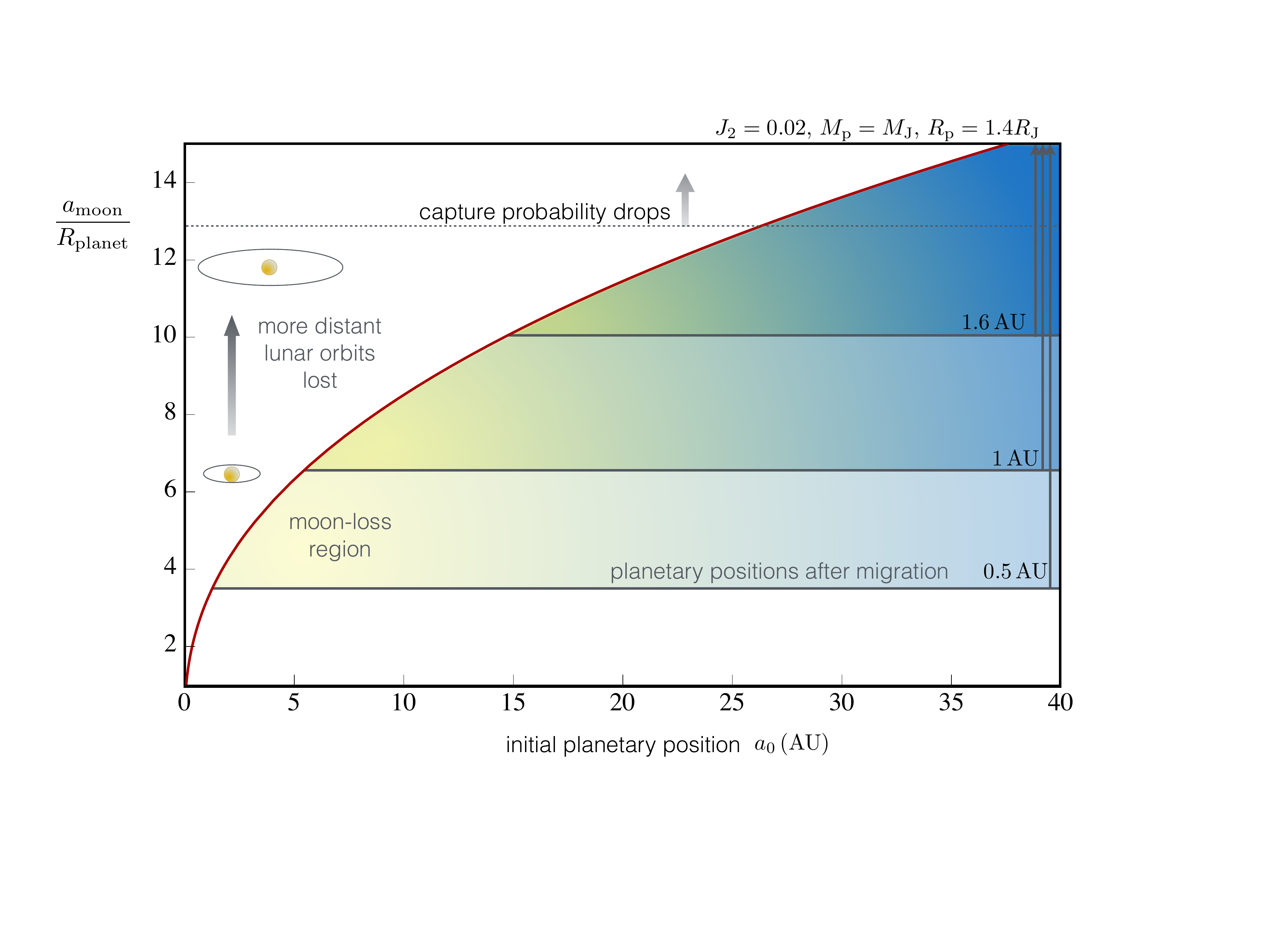}
\caption{The region of lunar orbit parameter space resulting in moon-loss. The shaded region between each horizontal line and the red curve illustrates lunar orbits lost when the host planet migrates from $a_0$ (horizontal axis) to the location denoted by the horizontal lines (1.6, 1.0 and 0.5\,AU). Requiring migration to occur sub-adiabatically (slowly) limits capture of orbits to $r\lesssim10$ (see text), corresponding to $a_0\sim15$\,AU. As an illustrative example, a Jupiter-like planet migrating from 15\,AU to 1\,AU may lose moons between $r\sim6.8-10$.}
\label{crash2}
\end{figure*}

The condition~(\ref{zone}) is general, but for clarity, in Figure~\ref{crash2} we present the extent of moon-removal appropriate to a Jupiter-mass planet around a Sun-like star. We display the specific regions of moon-loss for a planet currently found at $0.5$\,AU, 1\,AU and 1.4\,AU, as a function of $a_0$. It is clear that, provided the planet-moon system formed sufficiently far out, a significant extent of moon-space may be removed. For example, a hypothetical Jupiter, currently found at 1\,AU, could have lost a Europa-distanced moon ($r\approx6.8$) had the system originated at $\gtrsim5.3$\,AU.

It can be seen from Figure~\ref{crash2} that for each current planetary location (horizontal line), there exists a minimum initial location ($a_0=a_{\rm{crit}}$), below which no moons are lost (where the horizontal lines meet the red curve). This situation corresponds to when the migration extent is not sufficient to take any one lunar orbit all the way from circular to planet-crossing. We may approximate $a_{\rm{crit}}$ as a function of the final position $a_{\textrm{f}}$ by solving 
\begin{align}
r_{\rm{loss}}\bigg|_{a_{\textrm{f}}}=r_{\rm{Max}}\bigg|_{a_0=a_{\textrm{crit}}},
\end{align}
\noindent which yields the solution,
\begin{align}
a_{\textrm{crit}}=R_{\textrm{p}}\bigg[\bigg(\frac{2}{3J_2}\bigg)^{2}\bigg(\frac{M_\star}{M_{\textrm{p}}}\bigg)\,r_{\textrm{loss}}^7\bigg].
\end{align}

\subsection{Time required for moon-loss}
 
Having related the extent of planetary migration to a range of lost lunar orbits, we now consider how much time must pass in order to lose these moons and whether it may occur adiabatically. We mentioned above that the Type II migration timescale is expected to scale inversely with planetary mean motion. Accordingly, moons at larger distances from their planets (larger $r$), which are captured when their host planets cross more distant heliocentric radii, will be subject to much longer migration timescales than closer-in moons. Furthermore, imposing more rapid migration timescales would begin to impinge upon the adiabatic criterion~(\ref{criterion}). In what follows, we calculate the time taken for adiabatic moon-loss and compare it to the lifetime of a typical protoplanetary disk.

 Suppose the planetary semi-major axis evolves according to equation~(\ref{adot}), and that the migrationary timescale $\tau_m=\xi\, T_{\textrm{p}}$. In this case, we may calculate the time interval $\Delta t_e$ within which a given planet may migrate from some outer distance $a_0=a_{\textrm{res}}$ to an inner semi-major axis $a_{\textrm{c}}$. We do this through the solution of 
 \begin{align}
\frac{1}{a_{\textrm{p}}} \frac{d a_{\textrm{p}}}{dt}=- \frac{1}{2\pi\,\xi}\sqrt{\frac{GM_\star}{a_{\textrm{p}}^3}}
 \end{align}
whereby we obtain 
\begin{align}\label{traverse}
\Delta t_e=\frac{4\pi}{3}\xi \sqrt{\frac{a_{\textrm{res}}^3}{GM_\star}}\bigg[1-\bigg(\frac{2}{r}-\frac{1}{r^2}\bigg)^2\bigg],
\end{align}
where we have made use of relationship~(\ref{collision}), derived above, which states that 
\begin{align}
a_c=a_{\textrm{res}}\bigg(\frac{2}{r}-\frac{1}{r^2}\bigg)^{\frac{4}{3}}.
\end{align}

\begin{figure}
\centering
\includegraphics[trim=3cm 4.9cm 4cm 3cm, clip=true,width=1\columnwidth]{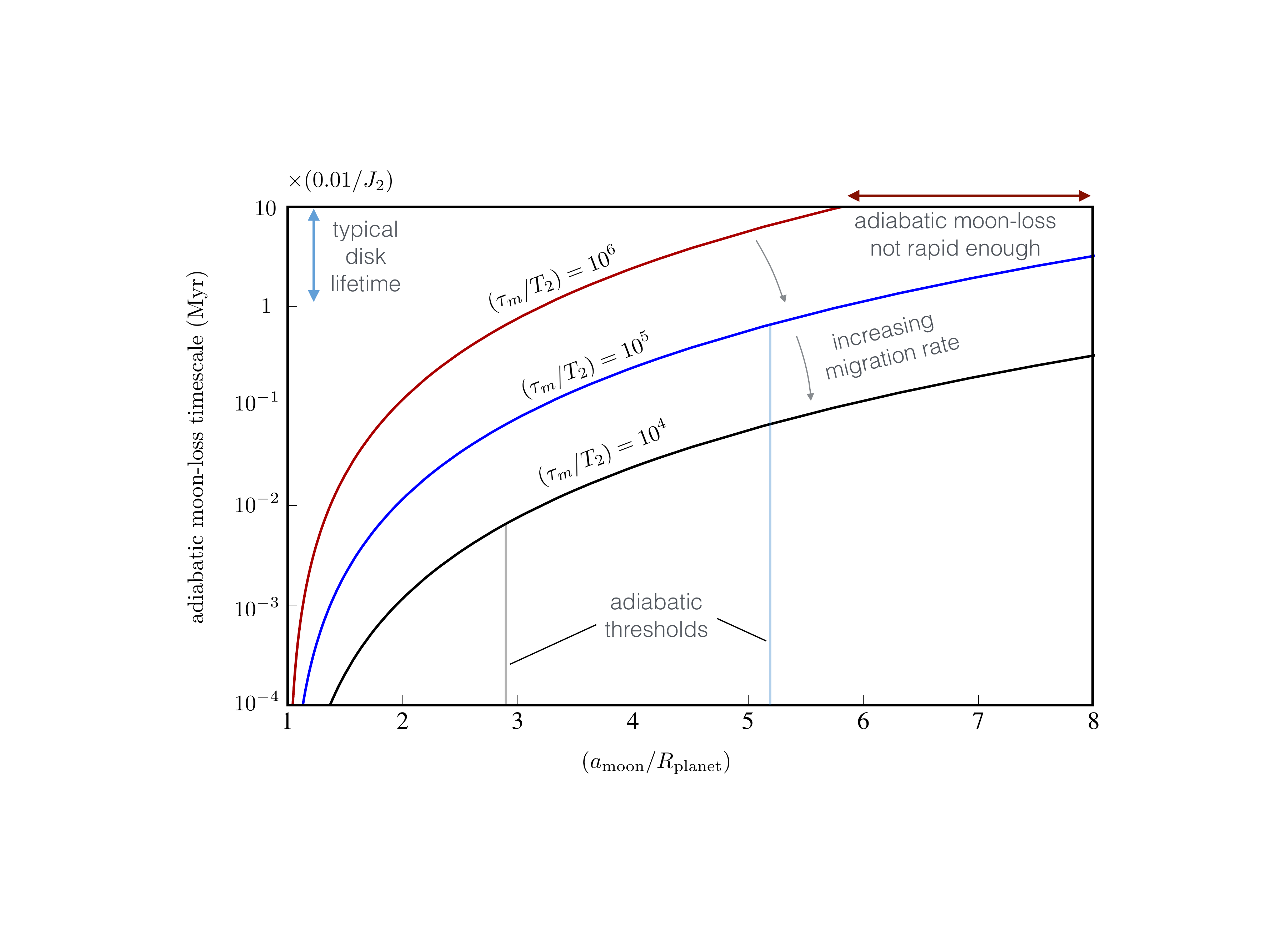}
\caption{Time taken to adiabatically lose a moon by way of the evection resonance. Capture into resonance occurs with certainty only below a threshold migration rate. The vertical lines indicate the most distant moon where the migration time $\tau_m/T_{\textrm{p}}=10^4$ (grey) and $\tau_m/T_{\textrm{p}}=10^5$ (blue) lead to adiabatic capture. A migration time of $10^4$ can traverse the resonant dynamics within a disk lifetime but is too rapid for adiabatic capture of moons beyond $\sim$\,2.9 planetary radii. Times scale as $J_2^{-1}$ and we have chosen $J_2=0.01$ for illustration. }
\label{MaxR}
\end{figure}

We illustrate the timescale $\Delta t_e$ in Figure~\ref{MaxR} for the cases $\xi=\{10^{\,4},\,10^5,\,10^6\}$ - all reasonable numbers given the current knowledge of Type II migration \citep{Armitage2010,Kley2012}. The critical planetocentric distance, below which moons may be adiabatically lost may be found by solving equation~(\ref{traverse}) for the value of $r$ such that $\Delta t_e$ equals some nominal disk lifetime $\tau_{\textrm{disk}}$. 

There are two competing effects at play. First, a planet must migrate slowly enough to capture its moon into resonance. Second, the planet must traverse a sufficient extent in semi-major axis, for its moon to crash into the planetary surface before the protoplanetary disk dissipates. What Figure~\ref{MaxR} suggests is that moons are unlikely to be adiabatically lost around Jupiter-mass planets if they lie further than $\sim10$ planetary radii away. In other words, if migration is slow enough for adiabatic capture, the disk dissipates before moon-loss at larger radii is complete. If migration is rapid, such as $\xi=10^{\,4}$, the planet can traverse the required distance in time, but only closer orbits ($r\sim3$) satisfy the adiabatic criterion, making capture of more distant moons rare. Here, we focus on Jupiter-like planets, but the range of lunar orbits over which adiabatic loss may occur expands for more massive planets, larger $J_2$ and less massive stars.  

 \begin{figure}
\centering
\includegraphics[trim=2cm 4.8cm 5.3cm 4cm, clip=true,width=1\columnwidth]{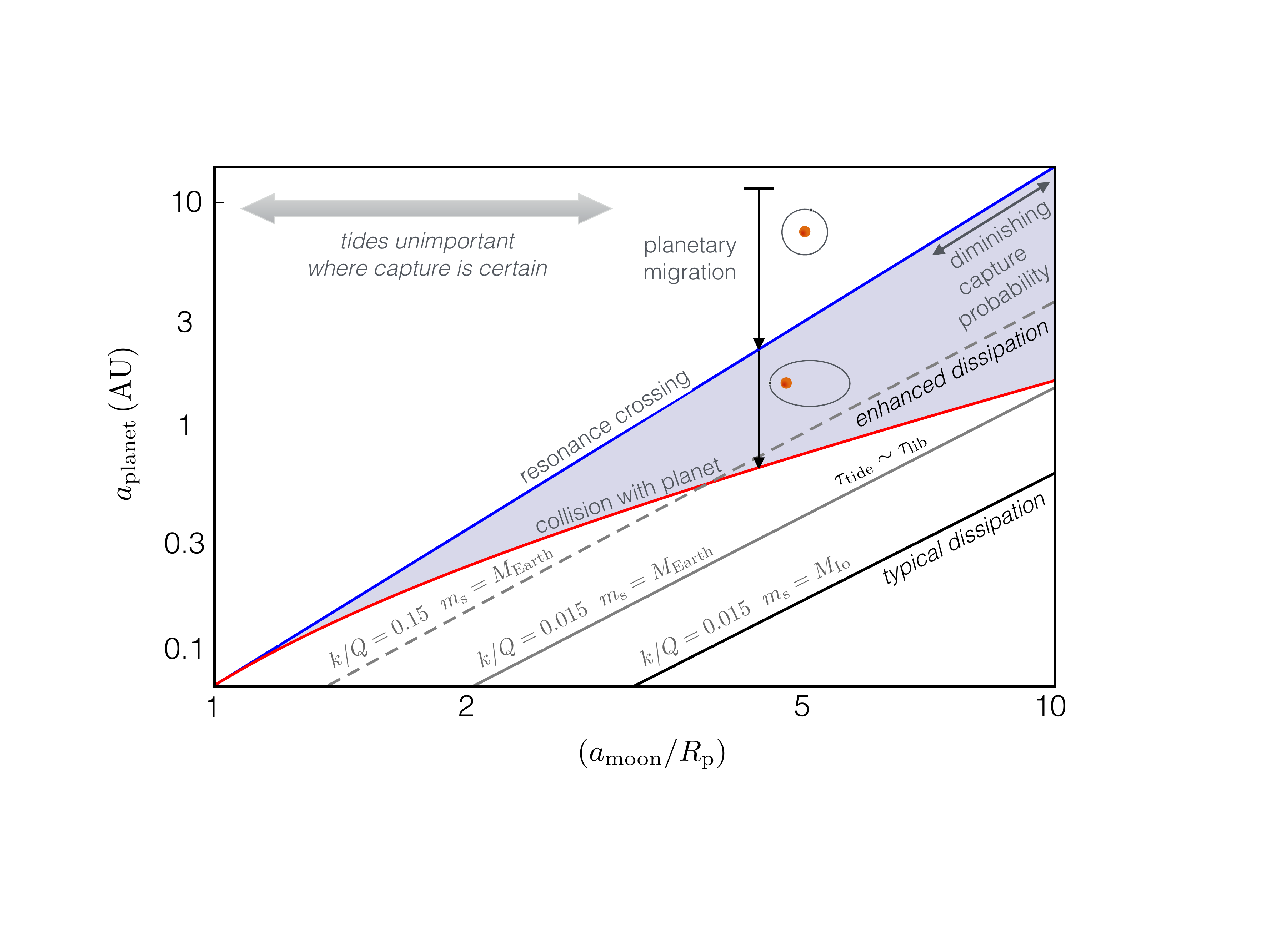}
\caption{An illustration of the required degree of migration for capture and loss of moons as a function of their planetocentric location. For clarity, we refer to the semi-major axes of the moon and planet as $a_{\textrm{moon}}$ and $a_{\textrm{planet}}$ (where `m' and `p' were used as subscripts in the text). The dashed, grey and black lines indicate the loci where tidal dissipation of eccentricity occurs as rapidly as libration. To the right of these lines, dissipation can break resonance. As can be seen, everywhere moons may be captured adiabatically, tides are unimportant, except when we artificially enhance the dissipation to roughly model the influence of continents and oceans in a  similar configuration to the modern day Earth.}
\label{TidesFigure}
\end{figure}

 \section{Dissipative effects}
 In the calculations presented above, we described the dynamics of a lunar orbit under the influence of purely gravitational forces. However, there exist two major sources of dissipation that may complicate the picture. First, moons around gas giants are thought to originate from within a circumplanetary disk of gas and dust \citep{Canup2002,Canup2006}. Analogously to planets embedded in circumstellar disks, moons are thought to interact with their disks such as to lead to inward migration of the moons in addition to a potential modulation of eccentricity. The second dominant source of dissipation is tidal planet-moon interactions, the strength of which is dependent upon both lunar eccentricity and semi-major axis \citep{Mignard1981,Hut1981}. We may generally suppose that dissipative influences will significantly alter the picture described above if the dissipation timescale is shorter than the libration timescale of the conservative Hamiltonian~(\ref{Ham}).

 \subsection{Tides}
 
 In this section, we discuss the influence of tidal dissipation upon the evection resonance. Specifically, the effect of the evection resonance is to increase eccentricity. Therefore, it is important to determine whether tidal damping of eccentricity will counteract its resonant growth. Recall that, in the conservative problem, we worked with the canonical variable $\tilde{\Gamma}=1-\sqrt{1-e^2}$, the rate of change of which is directly obtainable from Hamiltonian~(\ref{Ham}) by Hamilton's equation

\begin{align}
\dot{\tilde{\Gamma}}&=\frac{e}{(1-e^2)^{\frac{1}{2}}}\dot{e}=-\frac{\partial \mathcal{H}}{\partial \tilde{\gamma}_{\textrm{m}}}\nonumber\\
&=-n_{\textrm{p}}\,\frac{n_{\textrm{p}}}{n_{\textrm{m}}}\frac{15}{4}\tilde{\Gamma}(2-\tilde{\Gamma})\sin(2\tilde{\gamma}_{\textrm{m}})\nonumber\\
&=-n_{\textrm{p}}\,\frac{n_{\textrm{p}}}{n_{\textrm{m}}}\frac{15}{4}(1-\sqrt{1-e^2})(1+\sqrt{1-e^2})\sin{2\tilde{\gamma}_{\textrm{m}}},\nonumber\\
\end{align} 
 where we now suppose that $\sin(2\tilde{\gamma}_{\textrm{m}})\rightarrow 1$ because this corresponds to the maximum restoring torque that the conservative dynamics can apply. The tidal damping must overcome this eccentricity forcing if it is to break the system out of resonance. Therefore, we may write the conservative eccentricity growth as 
 
 \begin{align}
 \frac{de}{dt}\bigg|_{\textrm{grav}}\sim n_{\textrm{p}}\,\frac{n_{\textrm{p}}}{n_{\textrm{m}}}\frac{15}{4}\,e\,(1-e^2)^{\frac{1}{2}}.
 \end{align}
 
 The degree to which tidal eccentricity damping operates is somewhat uncertain, especially for general eccentricity. However, we obtain an approximate expression for the tidal damping by utilising the tidal formulae of \citet{Hut1981}. Specifically, we may approximate the tidal evolution of eccentricity by  
  \begin{align}\label{SatTide}
 \frac{de}{dt}\bigg|_{\textrm{tides}}=&-\frac{27k_{\textrm{m}}\,n_{\textrm{m}}}{2Q_{\textrm{m}}}\bigg(\frac{M_{\textrm{p}}}{m_{\textrm{s}}}\bigg)\bigg(\frac{R_{\textrm{s}}}{a_{\textrm{m}}}\bigg)^5e\,(1-e^2)^{-\frac{13}{2}}\nonumber\\
&\times \bigg[f_3(e^2)-\frac{11}{18}(1-e^2)^{\frac{3}{2}}f_4(e^2)\frac{\Omega_{\textrm{m}}}{n_{\textrm{m}}}\bigg]\nonumber\\
 \end{align}
 where, for $\Omega_{\textrm{m}}$, we consider the satellite to be in the equilibrium spin state ($\dot{\Omega}_{\textrm{m}}=0$). In Appendix~\ref{AppB}, we provide a brief derivation of the functional form of the equilibrium $\Omega_{\textrm{m}}$, which evaluates to 
  \begin{align}\label{Locked}
 \Omega_{\textrm{m}}=n_{\textrm{m}}\frac{f_2(e^2)}{(1-e^2)^{\frac{3}{2}}f_5(e^2)}.
 \end{align}

 In the above equations, $R_{\textrm{s}}$ is the satellite's physical radius and $m_{\textrm{s}}$ is its mass. In addition, we must specify the tidal love number $k_2$ and the quality factor $Q$, which are highly uncertain even in well-studied solar system bodies, such as the Galilean satellites, let alone hypothetical exomoons \citep{Lainey2009}. Accordingly, we choose three reasonable cases. First, we consider two moons with dissipative parameters appropriate for Io, with $k_2/Q\approx0.015$, but with one having the mass of Earth and the other the mass of Io. Owing to the dependence of the tidal damping upon satellite radius and mass, the Earth-mass moon will dissipate eccentricity more rapidly, assuming similar $k/Q$. As a third case, we note that a truly ``Earth-like" moon will come complete with oceans and continents, from which the vast majority of Earth's tidal dissipation stems \citep{Egbert2000}. Therefore, in the interest of completeness, we consider a scenario where the moon has the mass and radius of Earth, but tidal parameters ten times that of Io. The factor of ten is somewhat arbitrary and is taken simply to illustrate an extreme case. However, we note that the model proposed by \citet{Touma1998} required Earth's dissipation to be about 25 times weaker in the past to match the moon's current position, so an order of magnitude amplification is at least feasible.

 In Figure~\ref{TidesFigure}, we plot the locus of parameters where 
  \begin{align}
 \frac{de}{dt}\bigg|_{\textrm{tides}}= \frac{de}{dt}\bigg|_{\textrm{grav}}
 \end{align}
 for the three different cases described above. In general, tides act over too long of a timescale to break the resonance. However, where tides are artificially enhanced (the dotted line in Figure~\ref{TidesFigure}), moons residing beyond $r\sim3-4$ may be broken out of the resonance before destruction. Accordingly, the remote possibility exists that some habitable, Earth-like moons have been saved from annihilation by the very oceans and continents that make them habitable\footnote{This is somewhat of a fun speculation rather than a serious statement.}.

 \subsection{Influence of a Circumplanetary disk}
 
 Despite decades of work, the exact mechanisms governing turbulence, migration and planet formation within circumstellar disks remain elusive. Therefore, to claim a precise understanding of the analogous disks encircling young planets would be premature. However, the properties of moons around our own gas giants, Jupiter and Saturn, have helped guide sophisticated models of circumplanetary disks \citep{Canup2002,Canup2006,Martin2011}. In particular, moons are thought to undergo inward migration within such disks, a process we have thus far neglected. 
 
 A theory was put forward in \citet{Canup2006} to explain the conspicuously uniform mass ratio ($\sim 10^4$) between the masses of the planets Saturn, Jupiter and Uranus, and their respective satellite systems. The theory relies upon disk-driven migration (of the Type I variety) carrying previous generations of moons into the host planet, leaving only a surviving remnant whose formation time was similar to their migration time. If this picture is persistent across extrasolar giant planets, then the possibility exists that no one particular moon will ever stick around long enough for the evection resonance to remove it. However, with data limited to our own Solar System, it is not yet clear whether significant migration of moons does indeed occur during the epoch of planetary migration. 
 
 The treatment of migration within a circumplanetary context possesses several key differences from that within circumstellar disks. First, there are currently no observational constraints upon accretion rates within disks encircling planets. Indeed, once the planet has acquired its gaseous envelope in a runaway fashion, there is no strict requirement that the circumplanetary material accrete at all. In Appendix~\ref{AppC}, we provide a brief calculation of the steady-state disk mass arising from a maximum possible accretion rate of about one Jupiter mass per million years. For turbulence parameters typical of circumstellar disks, the total disk mass could be smaller than that of Io, but conversely, if turbulence was generated by the disk reaching a gravitationally unstable state, then the disk could easily be orders of magnitude more massive. Owing to such uncertainties, for the sake of this work, we simply mention that significant inward lunar migration would delay resonance-crossing, with the details left for case-by-case considerations. 
 
 If moon formation does indeed occur by way of a rapid creation and destruction of moons as they sequentially cascade into the host planet \citep{Canup2006}, then the evection resonance will only apply to the final generation of moons, when its efficacy depends upon how much more planetary migration occurs after this point in time. Similar arguments apply to disk-driven eccentricity damping. The evection resonance will only proceed once disk-driven eccentricity damping slows down, which ought to occur within a similar epoch to when satellite loss ceases and so we meet the same conclusion, that only the final generation of moons is subject to evection-induced moon loss.
 
 \section{Discussion}

 In this paper, we have identified a mechanism by which moons may be dynamically lost as their host planet undergoes Type II migration to shorter-period orbits. Specifically, inward migration increases the planetary orbital frequency until it becomes commensurate with the $J_2$-forced pericenter precession rate. Capture into this ``evection" resonance, followed by subsequent migration, drives the lunar orbit's eccentricity higher until the moon collides with the planet. We have shown that this mechanism is generally constrained to remove moons closer than about $\sim10$ planetary radii from their host planets. More distant moons may enter the resonance, but in their case, moon-loss is unlikely to occur within the typical lifetime of a protoplanetary disk. 

 \subsection{Determining migration extent}
 
 It has long been suspected that giant planets must form outside of the ice lines of their natal disks \citep{Pollack1996}, but exactly where they form is still an open question. An observational prediction of the mechanism proposed here is that moons should be more abundant at small and large planetocentric distances but rare at intermediate distances. The unoccupied region may be used to constrain where migration began, via Equation~(\ref{zone}). Such an inference has important implications for the formation pathway of giant planets. In particular, core-accretion is not thought to operate efficiently at larger heliocentric distances. \citep{Pollack1996,Armitage2011}. In these cooler regions (beyond about 20-40\,AU), gravitational instability and disk-fragmentation have been tentatively proposed, but are generally disfavored. 
 
 Disk fragmentation is occasionally invoked to explain very distant giant planets, such as those directly imaged in the system HR 8799 (heliocentric distances $\gtrsim40$\,AU; \citealt{Marois2008,Marois2010}). We have shown that moons cannot be lost adiabatically via the evection resonance when the planet originates at such distant radii. However, upon migration from $\sim10$\,AU, we do expect moons to be lost, depending upon where the planet ends its migration. Accordingly, we broadly expect evection-induced moon-loss to be more indicative of core accretion than of fragmentation, but caution that planets formed via the ``hot start" of disk fragmentation are likely to have enhanced radii than those arising from core accretion.

An important caveat is that we assume the planet and its moon form almost simultaneously. Were the moon to form after significant planetary migration, the region over which moons are lost could be significantly reduced. The extreme case thereof would be moons gravitationally caught subsequent to the disk-hosting phase, scenario we neglect. Proposed moon formation times are highly uncertain and mixed. For example, Callisto has been proposed to form before Jupiter's hypothesized inward migration \citep{Heller2015}. On the other hand, if moons are continuously being formed and lost within circumplanetary disks, no one single moon might be around long enough for loss via the evection resonance \citep{Canup2006}. Only once systems of exomoons are detected can we thoroughly test the competing hypotheses regarding moon formation and so for now we state that formation locations inferred from exolunar systems represent a lower bound on the actual location where the planet itself formed. 

In addition to determining the location at which a planet formed, we may also be able to constrain how much the planet has migrated outwards since the epoch of Type II migration. The three outer planets of our solar system are thought to have undergone significant post-nebular migration by way of planetesimal scattering \citep{Tsiganis2005}. Notwithstanding any influence such scattering has upon the moons, the observed inner edge of an excluded region may constrain where the planet resided at the end of Type II migration. A comparison to its observed location may extract some information regarding post-disk migration. It would be optimistic to expect particularly precise estimates via this method, but the general existence, or non-existence, of significant outward migration would help place the so-called ``Nice" model of our Solar System into its Galactic context. 

We have thus far neglected that, just as planets can move around after disk-dispersal, planetary tides can lead to significant evolution of satellite orbits. Indeed, it has even been proposed that the larger moons of planets within about 0.6\,AU of their host stars may be entirely lost (e.g. \citealt{Barnes2002}). In principle, it is possible to utilize tidal theory to infer where the moon once was around any given planet, but uncertainties are limited by the theory governing planet-moon tides, an active field of research even within our own solar system's satellites. With this in mind, we might suppose that younger planets are better targets, as they will have experienced less tidal evolution. 

In general, backing out the journey taken by a planet and its moon within their natal disk from the present lunar configuration is unlikely to become immensely precise for any one target. However, given a large enough sample of exomoon systems, we may begin to determine trends, or populations of planets with significantly different migrational histories that have thus far gone unnoticed, providing yet more impetus to continue searching for these objects.

\subsection{Implications for habitable moons}

Part of the motivation for this work was to determine whether Type II migration of giant planets may significantly reduce the number of moons occupying habitable zones. We have shown that these worlds are indeed subject to destruction through the evection resonance, but over a fairly restricted parameter space. Nevertheless, the mechanism is capable of reducing the population of habitable moons, except for the somewhat unlikely case that the satellite is as dissipative as Earth currently is, which is anomalously high even relative to Earth's geological history.

  As an illustration of the potential for habitable moon-loss, consider the horizontal line labelled ``1\,AU" in Figure~\ref{crash2}, appropriate to a Jupiter-mass planet currently situated at 1\,AU. Moons might be lost outside of $\sim6.3$ planetary radii upon migration from $\gtrsim5$\,AU. However, in order to lose moons beyond about 8 planetary radii, migration must take place within a disk lifetime, which corresponds to super-adiabatic motion for these parameters, making moon-loss unlikely. Accordingly, the moon-loss region is somewhat narrow in this specific case, but other cases may have significantly greater excluded regions. 

  \subsection{Additional considerations}
 
The assumptions adopted in our work inevitably leave room for future extensions to the framework. In particular, Jupiter's moons Io, Europa and Ganymede are locked in a 1:2:4 mean motion resonance, leaving open the question of how the picture changes if there are multiple moons around the migrating planet. Mean motion resonances are likely to quench the evection resonance as the apsidal recession driven through moon-moon interactions dominates over the evection-induced precession \citep{Murray1999,Morbidelli2002}. However, the picture is less clear when the moons are not locked in mutual resonances.

All discussion thus far has been with regard to destroying moons. However, suppose that the moon is caught into resonance, but the planet subsequently migrates only a small distance. The lunar orbit will thus be left eccentric but not planet-crossing. Further contraction of the planet after dissipation of the disk is analogous to inward migration within the disk because it reduces the coefficient of the $J_2$ term in the Hamiltonian. Accordingly, the moon would be pushed to yet higher eccentricities, potentially leading to a later collision with the planet. Alternatively, if sufficient contraction has occurred, the Roche Limit may lie outside the planetary radius and, in lieu of a collision, tidal forces would rip the moon apart to form a ring system. Indeed, tidal stripping from the icy, surface layers of a past moon has been invoked to explain the rings of Saturn \citep{Canup2006}, though it is unclear whether the evection resonance might have had a role in their formation. 

In tis work, we have introduced a novel mechanism for the removal of moons orbiting young, giant planets. Inward migration is expected to be almost ubiquitous in the formation of these planets, suggesting that the capture of moons into evection resonance is potentially a common process. We highlight that the resonance may be prevented by one of several mechanisms. Sufficiently rapid planetary migration can prevent capture and migration of the moon itself to shorter-period planetocentric orbits can delay or prevent resonance-crossing. Finally, the presence of other moons in the system, whether or not they exist in mean-motion resonance, can sometimes overpower evection. Such complications must be treated on a case-by-case basis as future exo-lunar detections emerge. \\
\textbf{Acknowledgements}
 CS acknowledges support from the NESSF student fellowship in Earth and Planetary Science. We acknowledge enlightening discussions with Dave Stevenson. We would also like to thank the referee, Matija \'Cuk, for his thoughtful report that greatly improved the manuscript. This research is based in part upon work supported by NSF grant AST
1517936.

\begin{appendices}
\section{Tidal equations}\label{AppB}
 
 In this section, we describe the equations used in determining tidal dissipation rates. Following \citet{Hut1981}, we adopt the following equations describing the tidal evolution of satellite spin-rate ($\Omega_{\textrm{s}}$) and eccentricity:
 \begin{align}
\frac{de}{dt}=&-27\frac{k_{\textrm{s}}}{T}q(1+q)\bigg(\frac{R_{\textrm{s}}}{a}\bigg)^8\frac{e}{(1-e^2)^{\frac{13}{2}}}\nonumber\\
&\times \bigg[f_3(e^2)-\frac{11}{18}(1-e^2)^{\frac{3}{2}}f_4(e^2)\frac{\Omega_{\textrm{s}}}{n_{\textrm{m}}}\bigg],\nonumber\\
\frac{d\Omega_{\textrm{s}}}{dt}=&3 \frac{k_{\textrm{s}}}{T}\frac{q^2}{I}\bigg(\frac{R_{\textrm{s}}}{a_{\textrm{m}}}\bigg)^6\frac{n_{\textrm{m}}}{(1-e^2)^6}\nonumber\\
 &\times \bigg[f_2(e^2)-(1-e^2)^{\frac{3}{2}}f_5(e^2)\frac{\Omega_{\textrm{s}}}{n_{\textrm{m}}}\bigg]\nonumber\\
\end{align}
where, adopting a constant-$Q$ framework,
\begin{align}
T\equiv\frac{R^3}{GM}2n_{\textrm{m}}Q.\nonumber\\
 \end{align}
 
 In the above equations, $q=M_{\textrm{p}}/m_{\textrm{s}}$ where $M_{\textrm{p}}$ is the mass of the perturbing body (the planet) and $m_{\textrm{s}}$ is the mass of the body upon which a tide is being raised (the satellite). Likewise, $R_{\textrm{s}}$ refers to the satellite's physical radius, $Q_{\textrm{s}}$ is the tidal quality parameter of the satellite and $k_{\textrm{s}}$ its tidal Love Number. The mass of the satellite $m_{\textrm{s}}$ is much smaller than the mass of the planet $M_{\textrm{p}}$, such that $q(1+q)\approx q^2$. The various functions $f_{\textrm{i}}$ are defined as
 
  \begin{align}
 f_2(e^2)&=1+\frac{15}{2}e^2+\frac{45}{8}e^4+\frac{5}{16}e^6\nonumber\\
 f_3(e^2)&=1+\frac{15}{4}e^2+\frac{15}{8}e^4+\frac{5}{64}e^6\nonumber\\
 f_4(e^2)&=1+\frac{3}{2}e^2+\frac{1}{8}e^4\nonumber\\
  f_5(e^2)&=1+3e^2+\frac{3}{8}e^4.\nonumber\\
 \end{align}
 
  The moment of inertia of the satellite $I$ is not important here because we suppose the satellite to be tidally locked ($\dot{\Omega}_{\textrm{s}}=0$) such that 
 
  \begin{align}
 \Omega_{\textrm{s}}=n\frac{f_2(e^2)}{(1-e^2)^{\frac{3}{2}}f_5(e^2)},
 \end{align}
 as claimed in the main text. When the appropriate substitutions are carried out, we arrive at equation~(\ref{Locked}).
\section{Circumplanetary Disk model}\label{AppC}
In what follows, we gain insight by considering a steady-state disk model, constrained to drive an accretion rate $\dot{M}$ lower than about 1 Jupiter mass every million years. If we adopt the \citet{Shakura1973} parameterization for effective viscosity, ignore any mass inflow and impose zero-torque inner boundary conditions \citep{Armitage2011}, the steady-state solution for surface density $\Sigma$ reads
 
 \begin{align}\label{sigma}
 \Sigma=\frac{\dot{M}}{3\pi \alpha}\frac{1}{\sqrt{G M_{\textrm{p}}a}}\bigg(\frac{h}{a}\bigg)^{-2}\bigg(1-\sqrt{\frac{R_{\textrm{p}}}{a}}\,\bigg),
 \end{align} 
 where $a$ is the planetocentric distance of the disk gas and $h$ is the pressure scale height of the circumplanetary disk. We may now estimate the mass of the disk by integrating from the planetary surface to some outer radius, $a_{\textrm{out}}$, which we estimate as the last non-crossing orbit inside the Hill radius \citep{Martin2011}, i.e., 
\be
r_{\textrm{out}} \approx 0.4\,r_H = \frac{2}{5}\,a_{\textrm{p}} \left( {M_{\textrm{p}} \over 3M_\star} \right)^{1/3} \,.
\ee
Using the above conditions, and taking the limit $a_{\textrm{out}}\gg R_{\textrm{p}}$, we obtain a disk with mass given by
 
 \begin{align}
M_{\textrm{disk}}\approx\frac{2 \dot{M}}{3\alpha \Omega_{\textrm{out}}}\bigg(\frac{h}{a}\bigg)^{-2}\frac{8}{45}\bigg(\frac{M_{\textrm{p}}}{3M_\star}\bigg)^{1/2},
 \end{align}
 where $\Omega_{\textrm{out}}$ is the orbital angular velocity of gas at the outer edge of the disk, which is related to the planet's mean motion via 
 
 \begin{align}
\Omega^2_{\textrm{out}}=3\bigg(\frac{5}{2}\bigg)^3 n^2_{\textrm{p}}.
 \end{align}
 
 Finally, we prescribe an accretion rate of $1\,M_{\textrm{J}}$\,myr$^{-1}$, such that $\dot{M}\sim M_{\textrm{J}}/\tau_{\textrm{acc}}$ with $\tau_{\textrm{acc}}=1$\,myr, leading to a disk mass
 
 \begin{align}
 \frac{M_{\textrm{disk}}}{M_{\textrm{p}}}\approx 10^{-3}\frac{T_{\textrm{p}}}{\tau_{\textrm{acc}}}\frac{1}{\alpha}\bigg(\frac{h/a}{0.2}\bigg)^{-2}\bigg(\frac{M_{\textrm{p}}/M_\star}{10^{-3}}\bigg)^{1/2}.
 \end{align}

Considering a planet at $1$\,AU around a Sun-like star, $T_{\textrm{p}}=1$\,year and so the disk mass we obtain is roughly 
 
  \begin{align}
 \frac{M_{\textrm{disk}}}{M_{\textrm{p}}}\approx 10^{-9}\frac{1}{\alpha}.
 \end{align}
 \noindent If $\alpha\sim10^{-3}$, then the inferred disk mass around a Jupiter-mass planet is significantly smaller than Io. However, the source of $\alpha$ is a mystery in these disks. Turbulence almost certainly commences once gravitational instability sets in but this requires $M_{\textrm{disk}}\sim (h/a)M_{\textrm{p}}$, suggesting very small $\alpha\sim10^{-8}$. 
 
 Such a diminutive $\alpha$ is not entirely unreasonable within the gravitationally-driven turbulence regime, provided disks have very long cooling times \citep{Gammie2001}. However, we are veering yet further into the unknown with these considerations and so we leave the details for future work. 
\end{appendices}

\begin{thebibliography}

\bibitem[Armitage(2010)]{Armitage2010}Armitage, P. J. (2010)., Cambridge University Press.

\bibitem[Armitage(2011)]{Armitage2011}Armitage, P. J. (2011), ARA\&A, 49, 195.

\bibitem[Barnes \& O'Brien(2002)]{Barnes2002}Barnes, J. W., \& O'Brien, D. P. (2002), ApJ, 575(2), 1087.

\bibitem[Batygin \& Adams(2013)]{Batygin2013} 
Batygin, K., \& Adams, F. C. 2013, ApJ, 778, 169

\bibitem[Borderies \& Goldreich(1984)]{Borderies1984}Borderies, N., \& Goldreich, P. (1984), Cel. Mech., 32(2), 127.

\bibitem[Canup \& Ward(2002)]{Canup2002}Canup, R. M., \& Ward, W. R. (2002), AJ, 124(6), 3404.

\bibitem[Canup \& Ward(2006)]{Canup2006}Canup, R. M., \& Ward, W. R. (2006), \textit{Nature}, 441(7095), 834.

\bibitem[Canup(2010)]{Canup2010}Canup, R. M. (2010), \textit{Nature}, 468(7326), 943.

\bibitem[Chandrasekhar(1957)]{Chandrasekhar1957}Chandrasekhar, S. (1957). An introduction to the study of stellar structure (Vol. 2). Courier Corporation.

\bibitem[Crida et al.(2006)]{Crida2006}Crida, A., Morbidelli, A., \& Masset, F. (2006), Icarus, 181(2), 587.

\bibitem[\'Cuk \& Stewart(2012)]{Cuk2012}\'Cuk, M., \& Stewart, S. T. (2012), \textit{Science}, 338(6110), 1047.

\bibitem[Danby(1992)]{Danby1992}Danby, J. (1992). Fundamentals of celestial mechanics. Richmond: Willman-Bell,| c1992, 2nd ed., 1.

\bibitem[Dawson \& Murray-Clay(2013)]{Dawson2013}Dawson, R. I., \& Murray-Clay, R. A. (2013),ApJ, 767(2), L24.

\bibitem[Domingos et al.(2006)]{Domingos2006}Domingos, R. C., Winter, O. C., \& Yokoyama, T. (2006), MNRAS, 373(3), 1227.

\bibitem[Duffell et al.(2014)]{Duffell2014}Duffell, P. C., Haiman, Z., MacFadyen, A. I., D'Orazio, D. J., \& Farris, B. D. (2014), ApJL, 792(1), L10.

\bibitem[Egbert \& Ray(2000)]{Egbert2000}Egbert, G. D., \& Ray, R. D. (2000), \textit{Nature}, 405(6788), 775.

\bibitem[Fortney \& Nettelmann(2010)]{Fortney2010}Fortney, J. J., \& Nettelmann, N. (2010), Space Science Rev., 152(1-4), 423.

\bibitem[Gammie(2001)]{Gammie2001}Gammie, C. F. (2001), ApJ, 553(1), 174. 

\bibitem[Haisch et al.(2001)]{Haisch2001}Haisch Jr, K. E., Lada, E. A., \& Lada, C. J. (2001), ApJL, 553(2), L153.

\bibitem[Hartmann et al.(1998)]{Hartmann1998}Hartmann, L., Calvet, N., Gullbring, E., \& D'Alessio, P. (1998), ApJ, 495(1), 385.

\bibitem[Heller et al.(2014)]{Heller2014}
Heller, R., Williams, D., Kipping, D., et al. 2014, 
AsBio, 14, 798

\bibitem[Heller et al.(2015)]{Heller2015}Heller, R., Marleau, G. D., \& Pudritz, R. E. (2015), A\&A, 579, L4.

\bibitem[Hut(1981)]{Hut1981}Hut, P. (1981), A\&A, 99, 126.

\bibitem[Iess et al.(2012)]{Iess2012}Iess, L., Jacobson, R. A., Ducci, M., Stevenson, D. J., Lunine, J. I., Armstrong, J. W., ... \& Tortora, P. (2012), \textit{Science}, 337(6093), 457.

\bibitem[Lainey et al.(2009)]{Lainey2009}Lainey, V., Arlot, J. E., Karatekin, \"O., \& Van Hoolst, T. (2009), \textit{Nature}, 459(7249), 957.

\bibitem[Lambrechts \& Johansen(2012)]{Lambrechts2012}Lambrechts, M., \& Johansen, A. (2012), A\&A, 544, A32.

\bibitem[Lissauer et al.(2011)]{Lissauer2011}Lissauer, J. J., Fabrycky, D. C., Ford, E. B., Borucki, W. J., Fressin, F., Marcy, G. W., ... \& Steffen, J. H. (2011), \textit{Nature}, 470(7332), 53.

\bibitem[King et al.(2007)]{King2007}King, A. R., Pringle, J. E., \& Livio, M. (2007), MNRAS, 376(4), 1740.

\bibitem[Kipping(2009)]{Kipping2009}Kipping, D. M. (2009), MNRAS, 392(1), 181. 

\bibitem[Kipping et al.(2009)]{Kipping2009b}Kipping, D. M., Fossey, S. J., \& Campanella, G. (2009), MNRAS, 400(1), 398.

\bibitem[Kipping et al.(2012)]{Kipping2012}
Kipping, D. M., Bakos, G. A., Buchhave, L., Nesvorny, D., \& 
Schmitt, A. 2012, ApJ, 750, 115 

\bibitem[Kipping et al.(2015)]{Kipping2015}
Kipping, D. M., Huang, X., Nesvorn\'y, D., Torres, G., Buchhave, L. A.,
Bakos, G. A., \& Schmitt, A. R. (2015), ApJ, 799, 14 

\bibitem[Kley \& Nelson(2012)]{Kley2012}Kley, W., \& Nelson, R. P. (2012). ARA\&A,, 50, 211.

\bibitem[Knutson et al.(2014)]{Knutson2014}Knutson, H. A., Fulton, B. J., Montet, B. T., Kao, M., Ngo, H., Howard, A. W., ... \& Muirhead, P. S. (2014), ApJ, 785(2), 126.

\bibitem[Machida et al.(2008)]{Machida2008}Machida, M. N., Kokubo, E., Inutsuka, S. I., \& Matsumoto, T. (2008), ApJ, 685(2), 1220.

\bibitem[Marley et al.(2007)]{Marley2007}Marley, M. S., Fortney, J. J., Hubickyj, O., Bodenheimer, P., \& Lissauer, J. J. (2007), ApJ, 655(1), 541.

\bibitem[Martin \& Lubow(2011)]{Martin2011}Martin, R. G., \& Lubow, S. H. (2011), MNRAS, 413(2), 1447.

\bibitem[Marois et al.(2008)]{Marois2008}Marois, C., Macintosh, B., Barman, T., Zuckerman, B., Song, I., Patience, J., ... \& Doyon, R. (2008), \textit{Science}, 322(5906), 1348.

\bibitem[Marois et al.(2010)]{Marois2010}Marois, C., Zuckerman, B., Konopacky, Q. M., Macintosh, B., \& Barman, T. (2010), \textit{Nature}, 468(7327), 1080.

\bibitem[Mayor \& Queloz(1995)]{Mayor1995}Mayor, M., \& Queloz, D. (1995), \textit{Nature}, 378(6555), 355.

\bibitem[Mignard(1981)]{Mignard1981}Mignard, F. (1981),The Moon and the Planets, 24(2), 189.

\bibitem[Morbidelli(2002)]{Morbidelli2002}Morbidelli, A. (2002). Modern celestial mechanics: aspects of solar system dynamics (Vol. 1).

\bibitem[Mosqueira et al.(2010)]{Mosqueira2010}Mosqueira, I., Estrada, P., \& Turrini, D. (2010), Space Science Rev., 153(1-4), 431.

\bibitem[Murray \& Dermott(1999)]{Murray1999} 
Murray, C. D., \& Dermott, S. F. 1999, Solar System Dynamics 
(Cambridge: Cambridge Univ. Press)

\bibitem[Nesvorny et al.(2003)]{Nesvorny2003}Nesvorn\'y, D., Alvarellos, J. L., Dones, L., \& Levison, H. F. (2003), AJ, 126(1), 398.

\bibitem[Quillen(2006)]{Quillen2006}Quillen, A. C. (2006), MNRAS, 365(4), 1367.

\bibitem[Pollack et al.(1996)]{Pollack1996}Pollack, J. B., Hubickyj, O., Bodenheimer, P., Lissauer, J. J., Podolak, M., \& Greenzweig, Y. (1996), Icarus, 124(1), 62.

\bibitem[Rowe et al.(2015)]{Rowe2015}Rowe, J. F., Coughlin, J. L., Antoci, V., Barclay, T., Batalha, N. M., Borucki, W. J., ... \& Zamudio, K. A. (2015), ApJS, 217, 16.

\bibitem[Shakura \& Sunyaev(1973)]{Shakura1973}Shakura, N. I., \& Sunyaev, R. A. (1973). A\&A, 24, 337.

\bibitem[Snellen et al.(2014)]{Snellen2014}Snellen, I. A., Brandl, B. R., de Kok, R. J., Brogi, M., Birkby, J., \& Schwarz, H. (2014), \textit{Nature}, 509(7498), 63.

\bibitem[Sterne(1939)]{Sterne1939}Sterne, T. E. (1939), MNRAS, 99, 451.

\bibitem[Stevenson(1982)]{Stevenson1982}Stevenson, D. J. (1982), Planetary and Space Science, 30(8), 755.

\bibitem[Takata \& Stevenson(1996)]{Takata1996}Takata, T., \& Stevenson, D. J. (1996), Icarus, 123(2), 404.

\bibitem[Tanaka et al.(2002)]{Tanaka2002}Tanaka, H., Takeuchi, T., \& Ward, W. R. (2002), ApJ, 565(2), 1257.

\bibitem[Touma \& Wisdom(1994)]{Touma1994}Touma, J., \& Wisdom, J. (1994). AJ, 108, 1943.

\bibitem[Touma \& Wisdom(1998)]{Touma1998} 
Touma, J., Wisdom, J. 1998, AJ, 115, 1653 

\bibitem[Tsiganis et al.(2005)]{Tsiganis2005}Tsiganis, K., Gomes, R., Morbidelli, A., \& Levison, H. F. (2005), Nature, 435(7041), 459.

\bibitem[Ward \& Hamilton(2004)]{Ward2004}Ward, W. R., \& Hamilton, D. P. (2004). Tilting saturn. I. Analytic model. The Astronomical Journal, 128(5), 2501.

\bibitem[Williams et al.(1997)]{Williams1997}Williams, D. M., Kasting, J. F., \& Wade, R. A. (1997), Nature, 385, 234.

\bibitem[Winn et al.(2010)]{Winn2010}Winn, J. N., Fabrycky, D., Albrecht, S., \& Johnson, J. A. (2010), ApJL, 718(2), L145.

\bibitem[Winn \& Fabrycky(2015)]{Winn2015}Winn, J., \& Fabrycky, D. C. (2015), ARA\&A, 53(1).

\bibitem[Yoder \& Kaula(1976)]{Yoder1976}Yoder, C. F., \& Kaula, W. M. (1976), Bull. of Am. Ast. Soc., (Vol. 8, p. 434).

\end{thebibliography}
\end{document}